\documentclass[11pt]{article}
\usepackage{amsmath,amsthm,latexsym,amssymb,amsfonts,epsfig,psfrag}

\makeatletter
\@addtoreset{equation}{section}
\makeatother

\def\be{\begin{equation}}
\def\ee{\end{equation}}
\def\ba{\begin{eqnarray}}
\def\ea{\end{eqnarray}}

\newcommand\nn{\nonumber}
\newcommand\q{\quad}
\def\Nl{{\mathchoice
{\setbox0=\hbox{$\displaystyle\rm N$}\hbox{\hbox to0pt
{\kern0.4\wd0\vrule height0.9\ht0\hss}\box0}}
{\setbox0=\hbox{$\textstyle\rm N$}\hbox{\hbox to0pt
{\kern0.4\wd0\vrule height0.9\ht0\hss}\box0}}
{\setbox0=\hbox{$\scriptstyle\rm N$}\hbox{\hbox to0pt
{\kern0.4\wd0\vrule height0.9\ht0\hss}\box0}}
{\setbox0=\hbox{$\scriptscriptstyle\rm N$}\hbox{\hbox to0pt
{\kern0.4\wd0\vrule height0.9\ht0\hss}\box0}}}}
\def\Zl{{\mathchoice
{\setbox0=\hbox{$\displaystyle\rm Z$}\hbox{\hbox to0pt
{\kern0.4\wd0\vrule height0.9\ht0\hss}\box0}}
{\setbox0=\hbox{$\textstyle\rm Z$}\hbox{\hbox to0pt
{\kern0.4\wd0\vrule height0.9\ht0\hss}\box0}}
{\setbox0=\hbox{$\scriptstyle\rm Z$}\hbox{\hbox to0pt
{\kern0.4\wd0\vrule height0.9\ht0\hss}\box0}}
{\setbox0=\hbox{$\scriptscriptstyle\rm Z$}\hbox{\hbox to0pt
{\kern0.4\wd0\vrule height0.9\ht0\hss}\box0}}}}
\def\Ql{{\mathchoice
{\setbox0=\hbox{$\displaystyle\rm Q$}\hbox{\hbox to0pt
{\kern0.4\wd0\vrule height0.9\ht0\hss}\box0}}
{\setbox0=\hbox{$\textstyle\rm Q$}\hbox{\hbox to0pt
{\kern0.4\wd0\vrule height0.9\ht0\hss}\box0}}
{\setbox0=\hbox{$\scriptstyle\rm Q$}\hbox{\hbox to0pt
{\kern0.4\wd0\vrule height0.9\ht0\hss}\box0}}
{\setbox0=\hbox{$\scriptscriptstyle\rm Q$}\hbox{\hbox to0pt
{\kern0.4\wd0\vrule height0.9\ht0\hss}\box0}}}}
\def\Rl{{\mathchoice
{\setbox0=\hbox{$\displaystyle\rm R$}\hbox{\hbox to0pt
{\kern0.4\wd0\vrule height0.9\ht0\hss}\box0}}
{\setbox0=\hbox{$\textstyle\rm R$}\hbox{\hbox to0pt
{\kern0.4\wd0\vrule height0.9\ht0\hss}\box0}}
{\setbox0=\hbox{$\scriptstyle\rm R$}\hbox{\hbox to0pt
{\kern0.4\wd0\vrule height0.9\ht0\hss}\box0}}
{\setbox0=\hbox{$\scriptscriptstyle\rm R$}\hbox{\hbox to0pt
{\kern0.4\wd0\vrule height0.9\ht0\hss}\box0}}}}
\def\Cl{{\mathchoice
{\setbox0=\hbox{$\displaystyle\rm C$}\hbox{\hbox to0pt
{\kern0.4\wd0\vrule height0.9\ht0\hss}\box0}}
{\setbox0=\hbox{$\textstyle\rm C$}\hbox{\hbox to0pt
{\kern0.4\wd0\vrule height0.9\ht0\hss}\box0}}
{\setbox0=\hbox{$\scriptstyle\rm C$}\hbox{\hbox to0pt
{\kern0.4\wd0\vrule height0.9\ht0\hss}\box0}}
{\setbox0=\hbox{$\scriptscriptstyle\rm C$}\hbox{\hbox to0pt
{\kern0.4\wd0\vrule height0.9\ht0\hss}\box0}}}}
\def\Hl{{\mathchoice
{\setbox0=\hbox{$\displaystyle\rm H$}\hbox{\hbox to0pt
{\kern0.4\wd0\vrule height0.9\ht0\hss}\box0}}
{\setbox0=\hbox{$\textstyle\rm H$}\hbox{\hbox to0pt
{\kern0.4\wd0\vrule height0.9\ht0\hss}\box0}}
{\setbox0=\hbox{$\scriptstyle\rm H$}\hbox{\hbox to0pt
{\kern0.4\wd0\vrule height0.9\ht0\hss}\box0}}
{\setbox0=\hbox{$\scriptscriptstyle\rm H$}\hbox{\hbox to0pt
{\kern0.4\wd0\vrule height0.9\ht0\hss}\box0}}}}
\def\Ol{{\mathchoice
{\setbox0=\hbox{$\displaystyle\rm O$}\hbox{\hbox to0pt
{\kern0.4\wd0\vrule height0.9\ht0\hss}\box0}}
{\setbox0=\hbox{$\textstyle\rm O$}\hbox{\hbox to0pt
{\kern0.4\wd0\vrule height0.9\ht0\hss}\box0}}
{\setbox0=\hbox{$\scriptstyle\rm O$}\hbox{\hbox to0pt
{\kern0.4\wd0\vrule height0.9\ht0\hss}\box0}}
{\setbox0=\hbox{$\scriptscriptstyle\rm O$}\hbox{\hbox to0pt
{\kern0.4\wd0\vrule height0.9\ht0\hss}\box0}}}}
\newcommand{\bd}{\mathbf d}

\title{(Broken) Gauge Symmetries and Constraints in Regge Calculus}
\author{Benjamin Bahr$^1$ and Bianca Dittrich$^2$\\
\small $^1$ DAMTP, University of Cambridge,\\
\small  Wilberforce Road, Cambridge CB3 0WA, UK \\
\small   $^2$ MPI f. Gravitational Physics, Albert Einstein Institute,\\
 \small Am M\"uhlenberg 1, D-14476 Potsdam, Germany\\
\small and\\
\small Institute for Theoretical Physics, Utrecht University,\\
\small  Leuvenlaan 4, NL-3584 CE Utrecht, The Netherlands
}

\date{\small Preprint AEI-2009-044}

\begin{document}

\maketitle

\begin{abstract}

We will examine the issue of diffeomorphism symmetry in simplicial
models of (quantum) gravity, in particular for Regge calculus. We
find that for a solution with curvature there do not exist exact
gauge symmetries on the discrete level. Furthermore we derive a
canonical formulation that exactly matches the dynamics and hence
symmetries of the covariant picture. In this canonical formulation
broken symmetries lead to the replacements of constraints by
so--called pseudo constraints. These considerations should be taken
into account in attempts to connect spin foam models, based on the
Regge action, with canonical loop quantum gravity, which aims at
implementing proper constraints.

We will argue that the long standing problem of finding a consistent
constraint algebra for discretized gravity theories is equivalent to
the problem of finding an action with exact diffeomorphism
symmetries. Finally we will analyze different limits in which the
pseudo constraints might turn into proper constraints. This could be
helpful to infer alternative discretization schemes in which the
symmetries are not broken.

\end{abstract}

\section{Introduction}\label{intro}

In quantizing a given theory its symmetries play a crucial role. The
question whether symmetries of the classical theory have also a
representation in the quantum theory can have a drastic influence on
the properties of the resulting quantum theory.

For general relativity the symmetry in question is diffeomorphism
invariance. As so far there is no satisfactory model of quantum
gravity yet, also the fate of diffeomorphism invariance in a quantum
theory of gravity is open.

Nevertheless a successful implementation of diffeomorphism
invariance into quantum gravity models could ensure the correct
semi-classical limit and moreover help to resolve quantization
ambiguities (see for instance \cite{flost}), that could otherwise
render the models unpredictive. It is therefore important to discuss
notions of diffeomorphism symmetries in the models at hand.

A particular class of models, for instance Regge quantum calculus
\cite{regge, rreview}, spin foam models \cite{spinfoams}, (causal)
dynamical triangulations \cite{cdt}, use discretizations of the
underlying spacetime manifold as a regulator in order to define the
specific model, i.e. a strategy how to perform the path integral. In
particular in spin foam models the Regge action appears in a
semi--classical limit \cite{conrady}. We will therefore concentrate
on the discussion of symmetries in Regge calculus.

The main question regarding diffeomorphism invariance for such
discretized models is whether a notion of exact diffeomorphism
invariance can be found for the discrete model, or whether exact
diffeomorphism invariance can arise only in a continuum limit
(alternatively in a sum over triangulations, that is all possible
ways of discretizations). A notion of exact diffeomorphism
invariance directly on the discrete level would simplify very much
the process of defining the theory, for instance in choosing the
path integral measure (which then has to be diffeomorphism
invariant). Furthermore, as we will show in this paper, with such a
notion it should be possible to find a canonical formulation for
discrete gravity models, with a closed, i.e. consistent constraint
algebra. This is the main problem for canonical or
Hamiltonian lattice gravity models \cite{canregge,lollc}.

In this work we will show, that for models based on the Regge action
(in 4d), diffeomorphism symmetry is generically broken. This is
contrary to expectations voiced in the literature \cite{hwgauge}.
Nevertheless there are many arguments that diffeomorphism invariance
will be restored in the continuum limit \cite{hartle1,moser,kasner},
and also our results will support this view.

These results do not exclude that discrete models with
diffeomorphism symmetry can be constructed. Indeed, the definition
of symmetry that we apply, depends crucially on the dynamics of the
model, as defined by the action, in this case the Regge action.
Other actions might exist, that exhibit exact diffeomorphism
symmetry. We will discuss such instances of different actions for
the same system with exact and broken symmetries respectively for 3d
Regge calculus with cosmological constant in section \ref{3dcan} and
toy models in section \ref{1d}.

A long standing problem is the construction of a consistent
discretized canonical model for gravity and a representation of
diffeomorphism in such a model, see for instance
\cite{waelbroeckzap,dittrichryan}. In a canonical formalism gauge
symmetries are reflected in constraints on the canonical data, that
also serve as generators for these gauge transformations. As
diffeomorphisms also include transformations of the time coordinate,
general relativity is a so called totally constrained system. That
is the Hamiltonian, the generator for the dynamics of the system, is
a combination of constraints. Hence the constraint algebra is of
central importance in order to have a consistent dynamics.

Often canonical lattice models are defined by discretizing the
constraints of the continuum theory. A typical problem in such
discretized models is that the constraint algebra is not closed.
This leads to severe problems for the quantization. According to the
Dirac program constraints resulting from gauge
symmetries have to be quantized and to be imposed onto the quantum
states. That is however only possible if the constraint algebra is
closed, or in other words anomaly free. Here the problem is already
on the classical level, i.e. one has to face classical anomalies.
Although different approaches exist to circumvent this problem
\cite{klauder,master,uniform}, it might be quite hard in these approaches to
keep classical and quantum anomalies and ambiguities under control.
This would be easier if we could construct discretized models with a
closed algebra. Here we will show that this problem can be seen to
be equivalent to finding an action, i.e. a covariant model with
exact diffeomorphism invariance. If the action displays exact gauge
invariance we will find constraints generating the gauge
transformations in the canonical formalism. For actions with broken
symmetries we will find a different picture: Instead of exact
constraints -- that is relations imposed by the dynamics that happen
to involve data of only one time step, we will find pseudo
constraints, that is dynamical relations which show a (weak)
dependence also on the data on the next time step. This dependence
can be interpreted as a dependence on the Lagrange multipliers lapse
and shift. This allows to solve the pseudo constraints for lapse and
shift. Hence gauge freedom is lost (as it is indeed broken on the
covariant level) and there are no constraints left on the canonical
data -- a picture that was advertised in the consistent
discretization program \cite{consistent}. We will however argue that
-- as one expects gauge symmetries to be restored in the continuum
limit -- it might be more promising to keep the pseudo constraints.
Indeed, data leading to solutions with a small discretization scale,
are concentrated on a `thickened constraint hypersurface'. In the
quantum theory the connection between covariant models with exact
gauge symmetry and canonical models with constraints is, that the
path integral acts as a projector onto the space of states
satisfying the constraints \cite{pro}. For systems with slightly
broken symmetries one would expect instead of an exact projector an
approximate implementation of the constraints, similar to a delta
function versus a Gaussian.

The advantage of the technique used in this work is that the
dynamics as defined by the covariant equations of motion and the
canonical time evolution equations coincide and hence also display
the same amount of gauge symmetry. These methods are in particular
important for attempts to establish a closer connection between spin
foam models and (canonical) loop quantum gravity. In the latter
dynamics is based on exact constraints whereas in the former we
expect diffeomorphism symmetry to be broken (as it is broken for the
Regge action and the Regge action appears in a semi-classical
analysis \cite{conrady} of current spin foam models). With this
different handling of symmetries by the two models the dynamics as
defined by these models very likely also differs - at least on the
discrete level.
\\[5pt]
In the next section we will discuss the notion of gauge symmetries
we are going to apply and how to test for the existence of these
symmetries. Next we discuss an evolution scheme for Regge calculus,
the so--called tent moves, which we are going to need for the
construction of a solution with curvature as well as for developing
a canonical formulation. We construct such solutions in section
\ref{4dsolution} and determine the eigenvalues of the Hessian for
these solutions. In section \ref{3dcan} we discuss a canonical
formulation for 3d discretized gravity with cosmological constant
using discrete actions with exact and with broken symmetries.
Furthermore we analyze limits in which the broken symmetries might
turn into exact ones. We will show in section \ref{relation} that if
we have exact gauge symmetries in the action then we will find
proper constraints on the discrete canonical data. Section \ref{1d}
provides another class of simple examples of discretized theories
with broken gauge symmetries. For these class of theories however
one can always define a discretization with exact gauge symmetries.
We will close with a discussion in section \ref{discussion}. The
appendices \ref{appamb} and \ref{formeln} contain a description of
two physically different solutions with the same boundary data and
formulas for geometrical quantities of simplices that we need in
section \ref{3dcan} respectively.

\section{Definition of gauge symmetry}\label{gauge}

In this section we will shortly describe the notion of
diffeomorphism symmetry that we are going to apply.

First of all we will consider continuously parametrized gauge
symmetries. (As discussed in \cite{bd08} there might also exist
notions which are completely based on discrete transformations, such
as a change of triangulation. We will comment shortly on a possible
relation to continuum symmetries in the discussion section.) In the
continuum such gauge symmetries lead to a continuous family of
solutions to the equations of motions (with fixed boundary values
for the variables), instead of just having one solution. We will
apply the same definition to discrete models. That is we will speak
of an (exact) gauge symmetry if the boundary value problem displays
non--uniqueness of solutions, moreover this non--uniqueness should
be parametrizable in a continuous way.

As solutions to the equations of motions are extrema of the actions,
this means that the action is constant in some directions exactly at
these extrema. As already discussed in \cite{bd08} it is important
to check that these constant directions also persist at the extrema
of the action, that is at solutions. (Away from solutions any
direction perpendicular to the gradient is a constant direction.)
This means that it is not sufficient to identify (possibly
configuration dependent) transformations that leave the action
invariant as these transformations might act trivially on solutions.
In this case we will still have uniqueness of solutions.

If there is a continuous parameter set of solutions then there exist
directions at these solutions in which the action is constant and
also the first derivatives of the action are constant (and equal to
zero). Hence in this case the Hessian, the matrix of second
derivatives, of the action, will have null eigenvectors. This
criterion is therefore a necessary condition for a gauge symmetry.

Note that different solutions might have gauge orbits of different
size. There might be theories with solutions with gauge symmetries
and solutions without gauge symmetries. As we will see this is the
case for 4d Regge calculus. There, flat solutions display gauge
symmetries \cite{rocekwilliams,fl,bdlfss}. Vertices of the
triangulation supporting this flat solution can be translated (in
four directions) without changing the flatness of the solutions. In
contrast, for solutions with curvature (and moreover without any
flat vertices\footnote{That is, all triangles adjacent to these
vertices have vanishing deficit angles.}) we do not find gauge
symmetries. Hence we have a mixture of exact gauge symmetries (for
flat vertices) and broken symmetries.

We will show explicitly that the criterion of vanishing eigenvalues
of the Hessian (evaluated at solutions) is violated for a 4d Regge
solution with curvature. But this example will also show that gauge
invariance is only slightly broken, hence we can speak of
approximate gauge invariance. Namely we will see that some of the
eigenvalues of the Hessian (per vertex) are very small as compared
to the rest of the eigenvalues. Moreover in approaching the flat
solution these eigenvalues go to zero (quadratically in the
curvature). Hence analyzing the eigenvalues of the Hessian is a
precise tool to discuss approximate gauge symmetries. This might be
helpful in order to construct actions with an exact gauge
invariance.

In section \ref{relation} we will furthermore see that the criterion
of null eigenvalues of the Hessian is related to the appearance (or
non--appearance) of constraints in a canonical formulation of the
theory.

\section{Tent moves}\label{sectent}

Tent moves are a way of defining a discrete time evolution for Regge
calculus, which has first been described in \cite{commi}, and used
for several works in Regge calculus \cite{kasner}. It is a way to
evolve a triangulated hypersurface locally, such that the
triangulation (that is the adjacency relations) of the resulting new
hypersurfaces does not change. These tent moves are a very
convenient tool to define a canonical formalism \cite{bd08} for
Regge calculus. Implementing some ideas from discrete numerical
integration \cite{marsden} or `consistent discretization'
\cite{consistent} the dynamics defined by the canonical and
covariant formulations will exactly coincide and therefore also the
gauge symmetries in these formulations.

Consider a $(d-1)$--dimensional triangulation $\Sigma$, which can be
thought of as a triangulated Cauchy hypersurface. Pick a vertex $v$
in the triangulation and define a new vertex $v^*$ lying in the
`future' of $v$, and connect both vertices with an edge. Denote all
other vertices in $\Sigma$ that $v$ is connected to by $1,\ldots,
n$. Connect also $v^*$ to the $1,\ldots, n$ by edges. Furthermore we
will have a simplex $v^*ij(k)$  (with vertices $v^*,i,j$ in 3d and
4d and $v^*,i,j,k$ in 4d) in the evolved hypersurface $\Sigma^*$ for
every simplex $vij(k)$ in $\Sigma$. Hence the triangulations of the
two Cauchy surfaces are the same. The evolution can be thought of as
gluing a certain piece of $d$--dimensional triangulation onto the
hypersurface. This $d$--dimensional triangulation consists of
simplices $vv^*ij(k)$ for every simplex $vij(k)$ in $\Sigma$ in
addition to simplices in the boundary coinciding with either
$\Sigma$, $\Sigma^*$ or both. The edge connecting $v$ and $v^*$ is
called the ``tent-pole''.


\begin{figure}[hbt!]\label{fig:TentMove5}
\begin{center}
    \psfrag{v}{$v$}
    \psfrag{vs}{$v*$}
    \psfrag{1}{$1$}
    \psfrag{2}{$2$}
    \psfrag{3}{$3$}
    \psfrag{4}{$4$}
    \psfrag{5}{$5$}
    \includegraphics[scale=0.6]{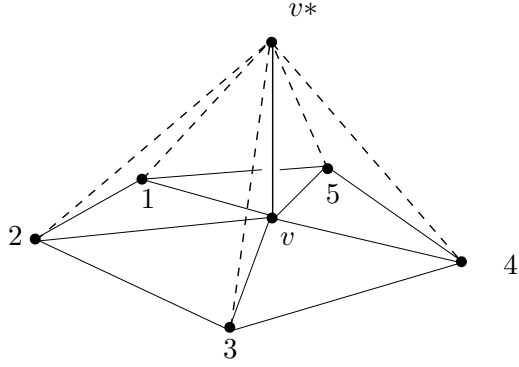}
    \end{center}
    \caption{\small A $5$-valent tent move at the vertex $v$.}
\end{figure}

\noindent

Note that each tent move can
be generated by a sequence of Pachner moves applied to the Cauchy surfaces. For $d=3$, an $n$-valent tent move is
the result of a $1-3$-move, followed by $(n-2)$ $2-2$ moves, and
finally a $3-1$ move. By applying tent moves to various vertices
after another,
one can
build up a large $d$-dimensional triangulation.

There are several advantages of this description:
\begin{itemize}
\item The evolution is local, in the sense that only the
triangulation containing the vertex $v$ (called the ``star of $v$'') is
evolved, while the rest of the triangulation remains untouched.
\item As a result, one can evolve a collection of vertices independently of each
other if neither of the vertices can be connected to any other of
the collection. This has in particular been implemented in numerical
applications \cite{kasner}.
\item The tent moves are particularly useful for an investigation of
Regge calculus in a canonical language, since all Cauchy
hypersurfaces that are produced in each step are isomorphic (as
simplicial complexes). Therefore, in each step the number of
canonical variables on the hypersurface remains unchanged. This is
not true in an arbitrary triangulation, which makes the canonical
analysis harder, and is the main reason why we consider tent moves
in this work. Furthermore for the analysis of a tent move we need to
consider only a small number of equations as opposed to a scheme in
which all vertices are evolved at once.

\item
The choice of which vertices of a hypersurface to evolve can be
understood as a discrete choice of lapse (to be either vanishing or
non--vanishing). Also, if $v_1$ and $v_2$ are vertices in the
initial triangulation that are connected by an edge, the two tent
moves applied to $v_1$ and $v_2$ do not commute. If one first
``evolves'' $v_1$ and then $v_2$, one obtains a different
($d$--dimensional) triangulation than first evolving $v_2$ and then
$v_1$. That might serve as a starting point for a definition of a
discrete notion for a hypersurface deformation algebra \cite{bd08}.
\end{itemize}

\subsection{Evolution equations for tent moves}

The Regge action for
a $d$--dimensional triangulation $T$ with boundary $\partial T$ and interior
$T^\circ:=T\backslash \partial T$ is given \cite{regge,bdry} by
\begin{eqnarray}\label{Gl:D-DimensionalReggeActionWithBoundary}
S_{\text{Regge}}\;=\;-\sum_{h\in
T^\circ}V_h\,\epsilon_h\;-\;\sum_{h\in\partial T}V_h\,\psi_h
+\lambda \sum_{\sigma \in T} V_\sigma
\end{eqnarray}

\noindent where the $h$ are the $d-2$ dimensional subsimplices
(sometimes called ``hinges'') of $T$ and $\sigma$ are the top--dimensional simplices of the triangulation.
$V_h$ and $V_\sigma$ denote the volume of the hinge $h$ and of the simplex $\sigma$ respectively. The deficit angles
$\epsilon_h$ and exterior angles $\psi_h$ are given by
\begin{eqnarray}\label{Gl:DefinitionDeficitAndExteriorAngles}
\epsilon_h\;&=&\;2\pi\;-\;\sum_{\sigma\subset
h}\theta_h^\sigma\qquad\text{for }h\in T^\circ\\[5pt]\nonumber
\psi_h\;&=&\;\pi\;-\;\sum_{\sigma\subset
h}\theta_h^\sigma\qquad\text{for }h\in\partial T,
\end{eqnarray}

\noindent and in both cases the sum ranges over all $d$--dimensional simplices $\sigma$ which contain $h$,
 and $\theta_h^\sigma$ is the
interior dihedral angle in the simplex $\sigma$ between the two
$(d-1)$--dimensional subsimplices that meet at $h\subset\sigma$.\\

Note that because of the boundary term in (\ref{Gl:D-DimensionalReggeActionWithBoundary}) the action is additive if we glue two pieces of triangulations together.

The equations of motion are obtained by varying the action
(\ref{Gl:D-DimensionalReggeActionWithBoundary}) with respect to the
lengths $l_e$ of the edges in $T^\circ$. The boundary term in
(\ref{Gl:D-DimensionalReggeActionWithBoundary}) ensures also that,
for all edges $e\in T^\circ$ the Regge equations read
\begin{eqnarray}\label{Gl:D-DimensionalReggeEquations}
-\sum_{h\supset e}\frac{\partial V_h}{\partial l_e}\,\epsilon_h\;+\;\lambda\sum_{\sigma \supset e} \frac{\partial V_\sigma}{\partial l_e}=\;0  \q .
\end{eqnarray}
The variation of the deficit angles appearing in the action vanishes because of the Schl\"afli identity, see appendix \ref{formeln}.

Consider a tent move for an $n$--valent vertex $v$ in the boundary
$\Sigma$ of a $d$--dimensional triangulation $T$. After the tent
move is performed we will have a new triangulation $T^*$ with
$(n+1)$ new inner edges, namely the edges $vi,\,i=1,\ldots n$
adjacent in $\Sigma$ to the vertex $v$ and the tent pole $vv^*$. We
will denote by $S_T,S_{T^*}$ and $S_*$ the action (with boundary
terms) of the original triangulation $T$, the new triangulation
$T^*$ and the piece of triangulation added in the tent move, so that
we have $S_{T^*}=S_T+S_*$. The equations for the new inner edges can
then be written as \ba\label{tentequ1}
\frac{\partial S_T}{\partial l_{vi}}+\frac{\partial S_*}{\partial l_{vi}}&=&0   \nn\\
\frac{\partial S_*}{\partial l_{vv^*}}&=&0  \q .
\ea

With the definitions
\begin{alignat}{2}\label{tentequ2}
p^-_{vi}&:=\frac{\partial S_T}{\partial l_{vi}} \q &\q\q  p^+_{vi}&:=-\frac{\partial S_*}{\partial l_{vi}}  \nn\\
p^-_{vv^*}&:=\frac{\partial S_T}{\partial l_{vv^*}} \q&\q\q  p^+_{vv*}&:=-\frac{\partial S_*}{\partial l_{vv^*}} \nn\\
p^{-}_{v^*i}&:=\frac{\partial S_*}{\partial l_{v^*i}} \q&\q\q  p^+_{v^*v^{**}}&:=-\frac{\partial S_*}{\partial l_{v^*v^{**}}}
\end{alignat}
the equations of motion (\ref{tentequ1}) are now given by
\ba\label{tentequ3} p^+_{vi}=p^-_{vi} \q\q \q
p^+_{vv^*}=p^-_{vv^*}=0 \ea that is the momenta $p^-$ and $p^+$
defined as derivatives of the actions associated to the two pieces
of the triangulation $T^*$ have to coincide. Therefore we will omit
the superindices $+,-$. With the second and third line in
(\ref{tentequ2}) we use the action of the added piece as a
generating function of first kind to define a canonical
transformation from the canonical variables
$(l_{vi},l_{vv^*},p_{vi},p_{vv^*})$ to a set of new canonical
variables $(l_{v^*i},l_{v^*v^{**}},p_{v^*i},p_{v^*v^{**}})$ (where
we introduced a fiducial vertex $v^{**}$ which can be thought of as
the vertex added in a second tent move). Here we see that the length
of the tent pole has a special status as its conjugated momentum is
constrained to vanish. Hence this variable is not fully dynamical.
This corresponds to an analogous result in the continuum where in
the canonical analysis the momenta conjugated to lapse and sift are
constrained to vanish.

Equations (\ref{tentequ2}) are just a reformulation of the equations
of motion in canonical language by using the action $S_*$ as a
generating function for a canonical transformation. This allows us
to obtain a canonical formalism which reflects exactly the dynamics
of the covariant formulation. Other attempts to define a canonical
framework for Regge calculus usually involve changing the dynamical
set up \cite{canregge}. Therefore gauge symmetries of the covariant
formulation might not be reflected properly in the canonical
formulation.

The advantage in the formulation used here is that the dynamics defined in the canonical formalism is the same as the covariant dynamics. Gauge symmetries of certain or all solutions should therefore have repercussions for the canonical formulation.

\section{Remarks on discrete ambiguities of solutions to the Regge equations}\label{ambig}

Given a boundary value problem the question arises whether the
solutions are unique. Gauge symmetries in the form discussed here
lead to a continuous family of non--unique solutions. In addition
there might be discrete ambiguities.

These also appear in Regge calculus, as for instance reported in
\cite{piran}. In our investigations we noticed ambiguities already
for the smallest boundary value problem, namely a triangulation with
only one inner edge. Such ambiguities are common to discretizations
of continuum theories. Typically there is only one solution that is
useful in a continuum limit whereas the others can be seen as
discretization artifacts. Nevertheless the question of discrete
ambiguities should be explored in more detail, as these might
influence the quantum theory.

Some of the ambiguities arise because of the following: If we
consider for instance a closed 4d ball with its 3d triangulated
boundary, then pieces of this 3d triangulation might stick inwardly
or outwardly. Similarly if we solve the tent pole equation (the
second equation in (\ref{tentequ1})) for the length of the tent pole
then typically there is one solution which is forward pointing, i.e.
a (bigger) tent is built on a smaller tent and another solution,
that is backward pointing, i.e. resulting in the tips of two tents
in opposite direction. Here we will always select the forward
pointing direction for the canonical analysis.

In the appendix \ref{appamb} we will discuss another kind of
ambiguity (arising by setting the prefactor $\frac{\partial
V_h}{\partial l_e}$ in the equations of motion
(\ref{Gl:D-DimensionalReggeEquations}) to zero). There we construct
a boundary value problem which allows for a flat and a curved
solution (with the same boundary data). The curved solution has
however very high curvature and one reason for its appearance seems
to be the highly symmetric situation. It can therefore considered to
be a discretization artifact.

\section{Construction of a Regge solution via tent moves}\label{4dsolution}

Here we will describe shortly how to find numerically a small Regge
solution with curvature using the tent moves. The evolution of
four--valent vertices should lead to flat solutions\footnote{Apart
from the discrete ambiguity described in the appendix.}: indeed such
solutions can be constructed by subdividing accordingly a flat
 4--simplex.

Therefore the simplest case to consider is the evolution
of a five--valent vertex. To this end we have to define the
three--dimensional triangulation around the vertex $v$, we want to
evolve, more concretely the three--dimensional star of $v$. As $v$
is five--valent we have five further vertices which we will denote
by $1,\ldots,5$. We will assume that we have six tetrahedra with
vertices \be\label{tent1} v124,\q v134,\q v234, \q  v125, \q , v135,
\q v235  \q . \ee Accordingly we will have nine triangles of the
form $vij$ with $i,j=1,\ldots 5$ in this triangulation, five edges
of the form $vi$ and nine edges of the form $ij$  (all possible
ordered combinations of $i,j\in \{1,\ldots 5\}$ with the exception
$45$).

\begin{figure}[ht]\label{fig:TentMove}
\begin{center}
    \psfrag{v}{$v$}
    \psfrag{1}{$1$}
    \psfrag{2}{$2$}
    \psfrag{3}{$3$}
    \psfrag{4}{$4$}
    \psfrag{5}{$5$}
    \includegraphics[scale=0.5]{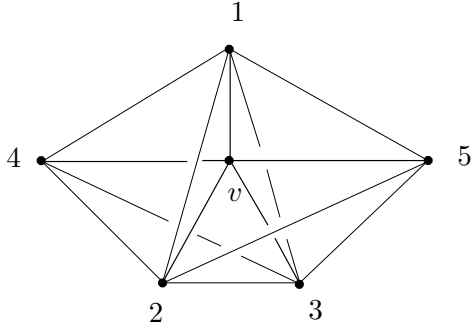}
    \end{center}
    \caption{\small The 3d boundary of the tent move triangulation.}
\end{figure}

To simplify the calculations even further we will assume that all
edges $ij$ have the same length $l_{ij}=s$. As the Regge vacuum
equations are invariant under a global rescaling  we will set $s=1$.
We will also assume that $l_{vi}=a$, $i=1,2,3$ are equal to each
other as well as $l_{v4}=l_{v5}=b$. Hence we have to deal with two
dynamical configuration variables $a$ and $b$. By $a^n,b^n$ we will
denote the values of these variables at the time step $n$. Together
with the lengths of the tent poles $t^n$ these will be also the free
variables in the action. (Varying the action with respect to all the
inner edge length and then looking for solutions with
$l_{vi}=a,l_{v\kappa}=b$ is equivalent to using this reduction in
the action and varying with respect to $a$,$b$ and the lengths of
the tent poles. In this sense we will consider a symmetry reduced
action.)

The 4--simplices involved in the tent move are all of the same type
$v^0v^1ij\kappa$ where $v^0,v^1$ denote the two vertices of the tent
pole (at time steps $n=0,n=1$ respectively), $i,j$ take values in
$1,2,3$ and $\kappa$ in $4,5$. We will denote by
\begin{itemize} \parskip -3pt
\item[~] $\theta^0_a,\, A^0_a$ the dihedral angle and the area of the triangle $v^0ij$,
\item[~] $\theta^0_b,\, A^0_b$ the dihedral angle and the area of the triangle $v^0i\kappa $,
\item[~] $\theta^a_t,\, A^a_t$ the dihedral angle and the area of the triangle $v^0v^1i $,
\item[~] $\theta^b_t,\, A^b_t$ the dihedral angle and the area of the triangle $v^0v^1\kappa $,
\item[~] $\theta^1_a,\, A^1_a$ the dihedral angle and the area of the triangle $v^1ij$,
\item[~] $\theta^1_b,\, A^1_b$ the dihedral angle and the area of the triangle $v^1i\kappa $ respectively .
\end{itemize}

The canonical equations of motion determining $a^1,b^1$ and $t^0$ given initial values $a^0,b^0$ and $p^0_a$ and $p^0_b$ are given by
\ba\label{tent2a}
0&=&3 \frac{\partial A^a_t}{\partial t^0} \left (2\pi  - 4 \theta^{a}_t  \right) +
2 \frac{\partial A^b_t}{\partial t^0} \left (2\pi  - 3 \theta^{b}_t \right)  \\
p^a_0 &=&  \frac{\partial A^0_a}{\partial a^0} \left (\pi  - 2 \theta^0_{a}  \right) +  \nn
  \frac{\partial A^0_b}{\partial a^0} \left (\pi  - 2 \theta^0_{b}  \right) +
 \frac{\partial A^a_t}{\partial a^0} \left (2\pi  - 4 \theta^a_t  \right) \q\q  \q\\
 p^b_0 &=&  3\frac{\partial A^0_b}{\partial b^0} \left (\pi  - 2 \theta^0_b  \right) +  \label{tent2b}
 \frac{\partial A^b_t}{\partial b^0} \left (2\pi  - 3 \theta^b_t \right)  \q\q .
\ea

The new momenta $p^a_1$ and $p^b_1$ are defined as
\ba\label{tent3}
p^a_1 &=& -\left(  \frac{\partial A^1_a}{\partial a^1} \left (\pi  - 2 \theta^1_{a}  \right) +
  \frac{\partial A^1_b}{\partial a^1} \left (\pi  - 2 \theta^1_{b}  \right) +
 \frac{\partial A^a_t}{\partial a^1} \left (2\pi  - 4 \theta^a_t  \right) \right) \q\q  \q \nn \\
 p^b_1 &=&  -\left(3\frac{\partial A^1_b}{\partial b^1} \left (\pi  - 2 \theta^1_b  \right) +
 \frac{\partial A^b_t}{\partial b^1} \left (2\pi  - 3 \theta^b_t \right)\right)  \q\q .
\ea

To construct a solution with an inner vertex $v^1$ we have to
consider at least two consecutive tent moves. We will proceed in the
following way. First we will start with some initial data
$(a^0,b^0,a^1,b^1)$ and use equation (\ref{tent2a}) to find the
length of the tent pole $t^0$. Given these five length we can
determine the momenta $(p^1_a,p^2_b)$  through equations
(\ref{tent3}). Now we have a full set of initial canonical data and
can use all three equations (\ref{tent2a},\ref{tent2b}) to find
$t^1,a^2,b^2$. We will end up with a Regge solution with inner
vertex $v^1$. In the end we have to evaluate the matrix of second
derivatives of the action with respect to $(a^1,b^1,t^0,t^1)$  on
this solution and determine its eigenvalues.

However if one attempts to solve the equations
(\ref{tent2a},\ref{tent2b}) for $t^1,a^2,b^2$ numerically one will
typically encounter the difficulty that standard numerical
(iterative) procedures do not converge. This is due to the fact that
there is at least an approximate gauge invariance (corresponding to
choosing the lapse at the vertex). So if we reformulate this problem
as finding the extrema of the action, we will have the difficulty
that there exist a direction in which the action is almost constant
and the extremum along this direction is difficult to locate. The
corresponding small eigenvalue of the Hessian leads to convergence
problems of the numerical procedures.

This difficulty is usually circumvented by for instance (gauge)
fixing the value for $t^1$ and ignoring one of the equations, for
instance (\ref{tent2a}) coming from the variation with respect to
$t^1$ \cite{kasner}. One then hopes that this equation is satisfied
up to a certain small error. This error is basically determined by
the value of the smallest eigenvalue in the Hessian.

To obtain a solution to a predetermined precision  $\delta$ we can
guess a value for $t^1$, solve the second and third equations for
$a^2$ and $b^2$ and then evaluate the right hand side of equation
(\ref{tent2a}). If this error $E(t_1)$ is larger than $\delta$ we
have to start again with another value of $t^1$. This procedure can
be systemized by starting with two values for $t^1$, say $A,B$ and
determining $E(A),E(B)$. If the signs of $E(A)$ and $E(B)$ are
different, than there is a zero in $[A,B]$ (assuming that $E$ is
continuous on this interval). This zero can be determined
iteratively to arbitrary precision.

We want to generate a set of solutions deviating slightly from the
flat solution. Hence we first construct a flat solution and then
introduce a deviation from the flat boundary data. Flat solutions
can be found by starting from a given data set $(a^0,b^0,t^0_f)$ and
considering the deficit angles at the triangles $v^0v^1i$,$i=1,2,3$
and $v^0v^1\kappa$,$\kappa=4,5$. Setting these to zero gives two
equations and we can find values $a^1_f(a^0,b^0,t_f^0)$ and
$b^1_f(a^0,b^0,t_f^0)$ satisfying these equations. Since we set
certain edge lengths equal to each other we have in this case only
one lapse degree of freedom, i.e. no shift degrees of freedom. This
means that we find one flat solutions for any given value of $t^0_f$
(and $a^0,b^0$).

Evaluating the Hesse matrix (second derivatives of the action with
respect to\\ $(a^1,b^1,t^0,t^1)$) on flat solutions we will find one
zero eigenvalue corresponding to the lapse degree of freedom. Hence
for a solution with curvature we expect either one vanishing
eigenvalue -- in case that the symmetry of the flat solutions
persists -- or at least one very small but non--vanishing eigenvalue
indicating that the symmetry is broken for curved solutions.

To obtain curved solutions we start with boundary data
$(a^0,b^0,a^1_f,b^1_f+x)$ deviating from the flat solution. We solve
numerically for $t^0,a^2,b^2$ and $t^1$ (with error terms of the
order $10^{-13}$).


\begin{figure}[ht]
\begin{minipage}[b]{0.5\linewidth}
\centering
\includegraphics[scale=0.7]{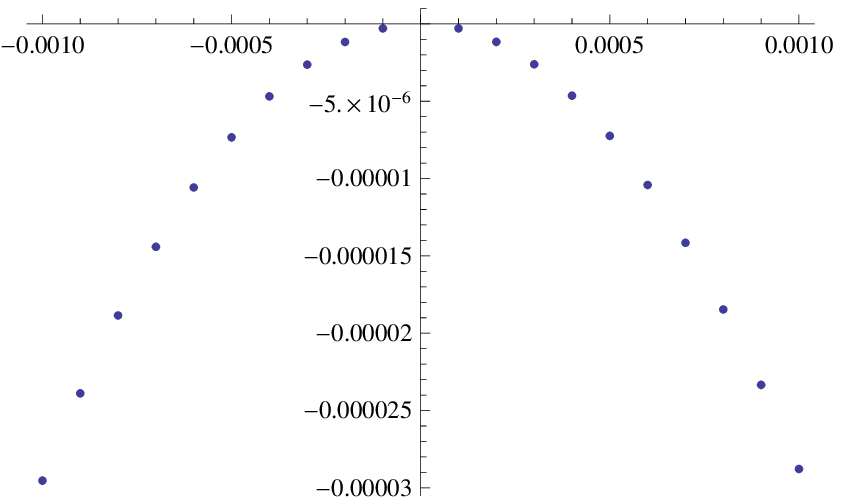}
\caption{The lowest eigenvalue of the Hessian as a function of the
deviation parameter $x$.} \label{fig:Hessian5v}
\end{minipage}
\hspace{0.3cm}
\begin{minipage}[b]{0.5\linewidth}
\centering
\includegraphics[scale=0.7]{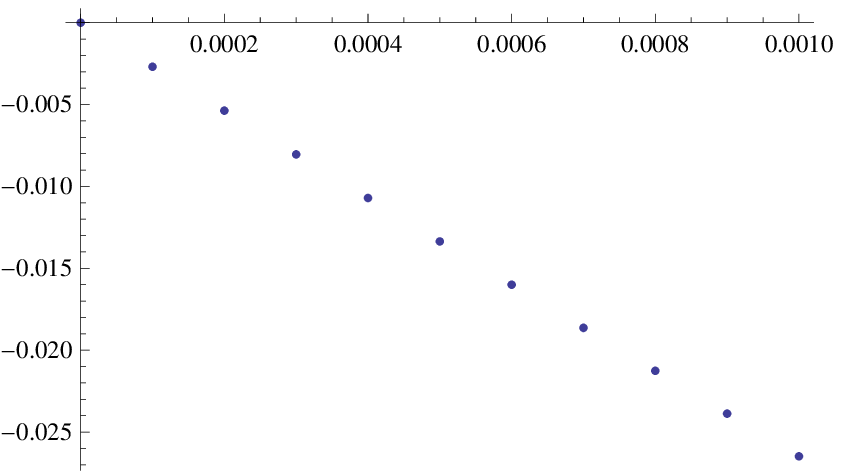}
\caption{The deficit angle at the triangle $v^0v^1i $ as a function
of the deviation parameter $x$.  } \label{eps5v}
\end{minipage}
\end{figure}

The plot in Figure \ref{fig:Hessian5v} is for initial data generated
as described above with $a^0=b^0=1$ and $a^1_f=1.16902,
b^1_f=1.13356$ corresponding to the choice $t^0_f=0.2$. As can be
seen from this plot, the lowest eigenvalue obtains non--zero values
for solutions with curvature. Compared to the other eigenvalues the
lowest eigenvalue is very small as can be seen for the set of data
given in Table \ref{eigenv}. In Figure \ref{eps5v} we plotted the
deficit angle at one of the inner triangles as a function of the
deviation parameter $x$ showing that we are indeed considering
solutions with non--vanishing curvature. Furthermore we see that the
lowest eigenvalue seems to grow quadratically with the curvature.

\begin{table}[ht]\begin{center}
\begin{tabular}{|c|c|c|c|c|}
\hline
x & eigenvalue 1 & eigenvalue 2& eigenvalue 3& eigenvalue 4\\
\hline
0.0001& 1678.52  & 570.127 & -394.796  & -2.91136$ \times 10^{-7}$\\
0.001& 1659.28& 562.471&-383.002 &-2.87768$\times 10^{-5}$ \\
0.01 & 1528.00&509.495&-306.545&-2.57318$\times 10^{-3}$\\
\hline\hline
\end{tabular}\end{center}
\caption{Eigenvalues of the Hessian as a function of the deviation
parameter $x$. The inner vertex has seven adjacent edges, but due to
the symmetry of the problem, these are associated to only four variables.}\label{eigenv}
\end{table}

This shows that for these vertices with curvature we do not have
symmetries which can be associated to translating the vertex. We
constructed also other examples (see appendix \ref{appamb} for one
other solution) where again for solutions with curvature we do not
find any vanishing eigenvalues of the Hessian. This does not exclude
that some curved solutions exist, which have vanishing eigenvalues -
however this seems to be a rather non--generic case.

Apart from the result that the symmetries in Regge calculus are
broken for vertices with curvature we want to point out that with
the methods presented here we can easily check to which extent the
symmetries are broken and how this scales with the curvature. The
example considered here indicates that the symmetry breaking grows
quadratically with the curvature. As we will see in section
\ref{3dcan} this might allow for a canonical formalism in which we
obtain proper constraints up to terms with a specific order in the
curvature.

\section{Canonical analysis for 3d Regge calculus with cosmological constant}\label{3dcan}

In this section we will investigate the consequences of broken and
exact gauge symmetries in the action for a canonical formalism.
Here we will discuss 3d Regge
calculus with and without a cosmological constant term as this example
provides us with both cases, one in which the symmetries are
exact and one in which they are broken. Moreover it allows us to
discuss limits in which the broken symmetries may become exact.
 The analysis of 4d Regge calculus (in which the symmetries are generically broken) will be postponed to a seperate paper \cite{hoehn}.

The solutions of 3d dimensional general relativity without a
cosmological constant are locally flat. Correspondingly the 3d Regge
equations just require that the deficit angles, which are attached
to the edges, should be vanishing. Hence every triangulation of a
locally flat space is a solution and moreover there is a
three--parameter gauge freedom attached to every inner vertex of the
triangulation. This gauge freedom corresponds to translating an
inner vertex such that the triangulation remains flat.

This gauge freedom vanishes if we consider a non--vanishing
cosmological constant and use the standard Regge action with an
added volume term \ba S= -\sum_{e \in T^\circ} l_e \epsilon_e
-\sum_{e\in \partial T} l_e \psi_e+\lambda \sum_{\tau \in T} V_\tau
\q \ea where $l_e,\epsilon_e$ are the length and the deficit angles
respectively associated to the edge $e$, $\psi_e$ is the exterior
angle at a boundary edge $e\in
\partial T$ and $V_\tau$ is the volume of a tetrahedron $\tau$  \cite{3dcons,bd08}.

\begin{figure}[ht]
\begin{minipage}[b]{0.5\linewidth}
\centering
\includegraphics[scale=0.75]{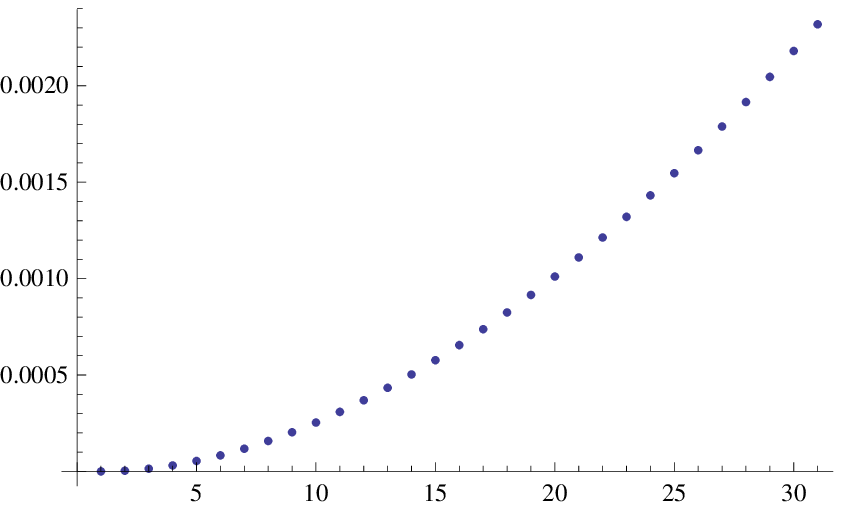}
\caption{The eigenvalue corresponding to translations in tent-pole direction as a function of $\lambda=0.01\times x$-axis label}
\label{3dhami}
\end{minipage}
\hspace{0.5cm}
\begin{minipage}[b]{0.5\linewidth}
\centering
\includegraphics[scale=0.75]{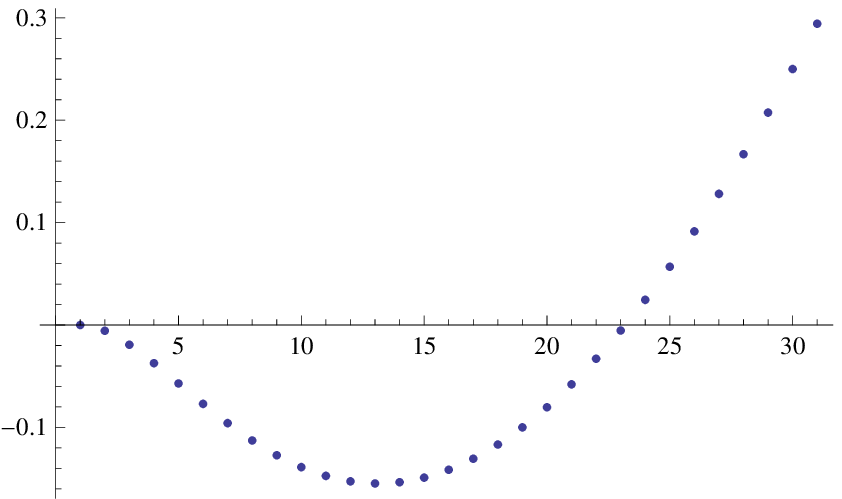}
\caption{  The eigenvalues corresponding to translations in
space-like direction as a function of $\lambda=0.01\times x$-axis
label              } \label{3ddiffeo}
\end{minipage}
\end{figure}

Indeed we can consider also in this case small 3d triangulations $T$
with an inner vertex and evaluate the Hessian on a solution. In this
case we used two consecutive tent moves at a three--valent vertex to
construct such a small 3d triangulation. The results show small but
non--vanishing eigenvalues corresponding to the approximate gauge
symmetries, see Figures \ref{3dhami} and \ref{3ddiffeo}. We
considered homogeneous solutions for which the three length
variables at the evolved vertex coincide. The initial values are
$l^0=1,l^1=1.8$ for the length variables at time step $0$ and $1$,
and $s=1$ for the non--dynamical edges not adjacent to either
$v^0,v^1,v^2$. Nevertheless we can consider the Hessian with
derivatives for all five variables (the three length variables and
the two tent pole variables) and evaluate this Hessian on the
homogeneous solution. Because of the homogeneous configuration there
are two small eigenvalues coinciding which correspond to
translations of the vertex in space-like directions (associated to
spatial diffeomorphism constraints), that is variations for which
the tent pole variables are constant. Then there is another small
eigenvalue corresponding to a translation in tent pole direction
(associated to a Hamiltonian constraint).

What makes the 3d case with cosmological constant interesting for us
is that  there is actually an action, defining a discretized
dynamics, in which these gauge symmetries are exact
\cite{bahrdittrich09b}. Starting with the standard 3d Regge action
we can therefore test different methods to construct an action with
exact gauge symmetries and see if we obtain the same result.

In this section we will start with the 3d standard Regge action and
perform a canonical analysis using tent moves. Since the gauge
symmetries are only approximate we will find that we do not obtain
exact constraints. Rather we have to deal with pseudo constraints
which depend on lapse and shift.

One could try to obtain exact constraints by considering a limit in
which the length of the tent pole (and therefore lapse and shift)
goes to zero. Indeed such a strategy works for discretized
reparametrization invariant systems as is explained in section
\ref{1d}. In this case however one cannot obtain useful results. The
reason is that gauge invariance is not being restored in this limit.
Note that here we keep the spatial discretization scale fixed and
send the discretization scale in time direction to zero. That
results in almost degenerate simplices. As is also argued in
\cite{moser} one cannot hope for a restoration of diffeomorphism
symmetry for such degenerate simplices.

On the other hand one can consider the limit in which both the
discretization scale in spatial and time directions are small. By
rescaling the edge length and the cosmological constant
appropriately this corresponds to small cosmological constant
$\lambda$. We will see that to first order in $\lambda$ the
constraints do not depend on lapse and shift. Moreover these
constraints correspond to the first order constraints obtained from
the alternative action with exact gauge symmetries.

\subsection{Canonical analysis of a tent move}

We will consider the tent move evolution of a three--valent vertex
$v$. We will denote the length of the edges between $v^n$, where the
superindex $n$ denotes the time step, and the adjacent vertices
$1,2,3$ by $l_i^n,\,i=1,2,3$. The length of the tent pole between
vertices $v^n$ and $v^{n+1}$ is $t^n$. Furthermore the
(non--dynamical) lengths of the edges $ij$ will be denoted by
$s_{ij}$. With $S_{(n,n+1)}$ we denote the action with boundary
terms for the piece of triangulation   that is added by performing a
tent move. In this case the added piece are three tetrahedra joined
at the tent pole.

\begin{figure}[ht]
\begin{center}
    \psfrag{v}{$v^n$}
    \psfrag{vs}{$v^{n+1}$}
    \psfrag{1}{$1$}
    \psfrag{2}{$2$}
    \psfrag{3}{$3$}
    \includegraphics[scale=0.4]{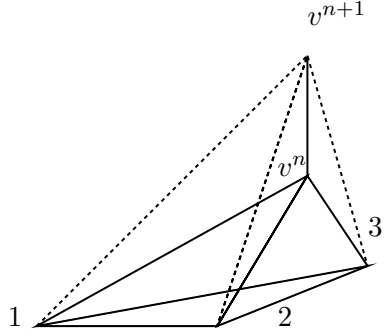}
    \end{center}
    \caption{\small A tent move for a three--valent vertex.}\label{3dtentpic}
\end{figure}

According to (\ref{tentequ2},\ref{tentequ3}) the canonical time evolution equations are given by
\ba\label{3d1}
p^n_i&=&-\frac{\partial S_{(n,n+1)}}{\partial l_i^n} = \psi^{n+}_{i}-\lambda \frac{\partial V_{(n,n+1)}}{\partial l_i^n}\nn\\
p^n_t&=&-\frac{\partial S_{(n,n+1)}}{\partial t^n} = \epsilon^{n}_t-\lambda \frac{\partial V_{(n,n+1)}}{\partial t^n}\nn\\
p^{n+1}_i &=&\,\,\frac{\partial S_{(n,n+1)}}{\partial l_i^{n+1}}\,\,=-\psi^{(n+1)-}_{i}+\lambda \frac{\partial V_{(n,n+1)}}{\partial l_i^{n+1}}\nn\\
p^{n+1}_t&=&\,\,\frac{\partial S_{(n,n+1)}}{\partial t^{n+1}} \,\,=0
\q . \ea Here $\psi^{n+}_i$ is the exterior angle at the edge $v^ni$
of the triangulation between time steps $n$ and $(n+1)$ whereas we
denoted with $\psi^{(n+1)-}$ the exterior angle at the edge
$v^{n+1}i$ of the same piece of triangulation. $\epsilon^n_t$ is the
deficit angle at the tent pole $v^nv^{n+1}$ and $V_{(n,n+1)}$ the
volume of the triangulation between the time steps $n$ and $(n+1)$.

Because of the last equation in (\ref{3d1}) we have that $p^n_t=0$
for all time steps $n$. We want to solve the second equation
$p^n_t=0$ for $t^n=t^n(l_i^n,l_i^{n+1})$ and use this solution in
the first equation. For $\lambda\neq 0$ this equation however
involves sums of trigonometric functions and cannot be solved
explicitly. Nevertheless it is possible to make an ansatz
$t^n={}^{0}t^n+\lambda\,\,{}^{1}t^n+\ldots$ and solve this equation
order by order in $\lambda$.

The zeroth order gives the dynamics for vanishing cosmological
constant, that is solutions are locally flat and all deficit angles
have to vanish. The solution ${}^0t^n$ to the second equation can be
easily constructed by geometrical considerations, it is the length
of the edge between the two tips of a double pyramid with triangular
base. (In general there are two solutions, one in which the tips of
the pyramid are in opposite direction and one in which the tips are
in the same direction. We will use the latter case.) Using this
solution in the first equation in (\ref{3d1}) (with $\lambda=0$) one
will find that the dependency on the length $l_i^{n+1}$ drops out
and that the momenta are constrained to be \ba\label{3d2}
p^n_i=-\pi+\theta_{v^ni}(l_i^n,s_{ij})  \q \ea where with
$\theta_{v^ni}(l_i^n,s_{ij})$ we denote the dihedral angle at the
edge $v^ni$ of a tetrahedron with edge lengths $l_i^n$ and $s_{ij}$
and with vertices $v^{n},1,2,3$.. Indeed since the dynamics of the
theory just results in flat space the intrinsic and extrinsic
geometry of the hypersurface resulting from the tent move evolution
will be just the same as that of the  boundary of a flat tetrahedron
embedded in flat 3d space. The constraints (\ref{3d2}) just express
this geometrical fact.

If we instead use the solution ${}^0t^n=\,{}^0t^n(l_i^n,l_i^{n+1})$
in the third equation of (\ref{3d1}), which defines the momenta
$p_i^{n+1}$ at the time step $(n+1)$ we will find that the
dependence on the length variables $l_i^n$ drops out and that the
constraints are preserved by time evolution\footnote{This can be
seen from the fact that $S_{(n,n+1)}$ is symmetric under exchange of
$l_i^n$ and $l_i^{n+1}$. The different signs in the definition of
momenta in (\ref{3d1}) are absorbed by the property of the momenta
to change sign for evolution in `forward' or `backward' direction.
See also the discussion for the momenta to first order in
$\lambda$.} \ba\label{3d3}
p^{n+1}_i=-\pi+\theta_{v^{n+1}i}(l_i^{n+1},s_{ij})  \q . \ea

To summarize, for $\lambda=0$, the canonical evolution equations
(\ref{3d1}) do not determine uniquely the evolution for the lengths
$l_i^{n+1}$ and the momenta $p_i^{n+1}$, they rather result in three
constraints determining the momenta $p^n_i$ and $p^{n+1}_i$ as
function of the lengths $l_i^n$ and $l_i^{n+1}$ respectively.
Therefore given three length $l_i^n$ the momenta $p^{n}_i$ are
determined by the constraints, whereas the three length $l_i^{n+1}$
at the next time step can be chosen freely (respecting generalized
triangle inequalities). This corresponds to a free choice of lapse
and shift at the vertex $v^n$. The set of six length variables
$l_i^n,l_i^{n+1}$ determines the length of the tent pole $t^n$.

Moreover one can check explicitly that the constraints (\ref{3d2})
are first class, even Abelian (see appendix \ref{formeln}). See also
\cite{dittrichfreidel}, where it is shown that taking into account
constraints based at different vertices we obtain a first class
algebra. The constraints are the infinitesimal generators for time
evolution -- the free choice of the three Lagrange multipliers
corresponds to the gauge choice of the three edge lengths
$l^{n+1}_n$ in the equations (\ref{3d1}). The choice of only one of
the three Lagrange multipliers $N_j=\varepsilon$ for the constraints
$C_i=p_i^n+\pi-\theta_{v^ni}(l_i^n,s_{ij})$ to be non--vanishing
corresponds to the choice $l_{k\neq j}^{n+1}=l^n_{k\neq j},\,
l_j^{n+1}=l_j+\varepsilon$ for the lengths at the next
(infinitesimal) time step.

If we switch to a non--vanishing cosmological constant the solutions
of the Regge equations do not display the same gauge freedom. This
has consequences for the canonical analysis.

The difference to the flat case is that if we solve (numerically)
the second equation of (\ref{3d1}) for the length of the tent pole
and use this solution $t^n(l_i^n,l_i^{n+1})$ in the first equation
the dependence of the momenta $p_i^n$ on the lengths $l_i^{n+1}$
will not drop out. Hence we can solve these equations for the
lengths $l_i^{n+1}$ at the next time step as a function of the
lengths $l_i^n$ and momenta $p_i^n$ at time step $n$. Using the last
equation in (\ref{3d1}) we can then determine also the momenta
$p_i^{n+1}$ at the next time step $(n+1)$. That is given some
initial data $l_i^0,p_i^n$ the evolution is unique for all other
time steps (ignoring discrete non--uniqueness and assuming the
existence of a solution).

\begin{figure}[hbt!]
\begin{center}
  \includegraphics[scale=1]{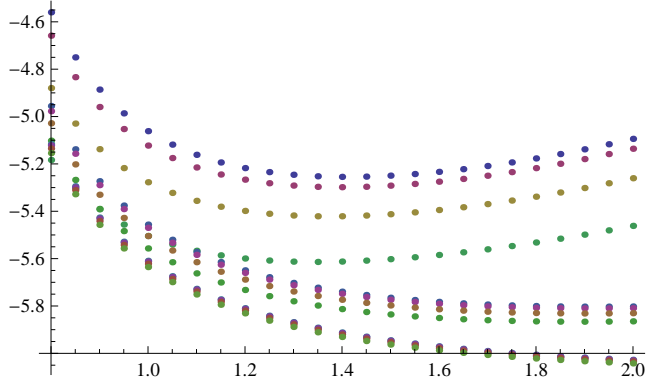}
    \end{center}
    \caption{\small The momentum $p^0$ as a function of the length $l^0$ for $\lambda=0.6,1.2,2.4$ and $l^1=l^0+0.01,0.4,0.8,1.2$.}\label{fig:3dsmall}
\end{figure}

Note however that the dependence of the momenta
$p^n_i(l_j^n,l_j^{n+1})$ on $l^{n+1}_j$ is very weak (for small
$\lambda$), see Figure \ref{fig:3dsmall}. Indeed as we will see
later, if we expand $p^n_i(l_j^n,l_j^{n+1})$ in $\lambda$ only the
second and higher order terms will depend on $l_j^{n+1}$.

Figure \ref{fig:3dsmall} shows the functions
$p^n_i(l_j^n,l_j^{n+1})$ evaluated on a homogeneous configuration
where $l_j^n=l^0$ and $l_j^{n+1}=l^1$ as well as $s_{ij}=1$. We
plotted the momenta as functions of $l^0$, but for different values
of $l^1$ ranging from $l^0+0.01$ to $l^0+1.2$. This is a wide range
considering that $l_0$ varies between $0.8$ and $2$. Moreover we
used $\lambda=0.6,1.2$ and $\lambda=2.4$. The four plots at the top
in Figure \ref{fig:3dsmall} are for the highest value of the
cosmological constant $\lambda$ and show quite a variation with
$l^1$. For $\lambda=1.2$ the variation is much smaller (the family
of plots below the four top plots). The plots for $l^1=l^2+0.01$ and
$l^1=l^0+0.4$ are almost above each other. Choosing canonical data
outside of the region traced out by the plots would lead to an
evolution with large edge length or would not allow for any (real)
solutions. Finally for $\lambda=0.6$ all the four plots are almost
above each other.

That is although we can apparently freely choose initial data
$l_i^0,p_i^0$ a corresponding solution might not exist or lead to
very large edge lengths. If one wants for instance the lengths
$l_i^1$ at the following time step to be in rather small intervals
around $l_i^0$ (such that the length $t^0$ is small), the momenta
$p_i^0$ will be also quite restricted to a small interval. For the
flat case we argued that the gauge choice of the $l_i^{n+1}$
corresponds to a choice of the Lagrange multipliers lapse and shift.
If we consider a non--vanishing cosmological constant we can
interpret the functions \ba\label{3d4}
C_i^\lambda=p^n_i-p^n_i(l_j^n,l_j^{n+1}) \ea as lapse and shift
dependent constraints. For lapse and shift in small intervals we
will obtain instead of one constraint hypersurface of infintesimal
width a family of hypersurfaces labeled by lapse and shift or in
other words a thickened constraint hypersurface of finite width. In
the limit of vanishing cosmological constant (but keeping the lapse
and shift intervals fixed) the width of this thickened constraint
hypersurface converges to zero.

Next we will discuss limits in which one may regain proper
constraints. To obtain such constraints for finite $\lambda$ one
might try to get rid of the $l^{n+1}_i$--dependence of the pseudo
constraint $C_i^\lambda$ in (\ref{3d4}) by considering the limit
$l_i^{n+1}\rightarrow l_i^n$ (in case it is well defined). It
corresponds to a limit in which the discretization scale in time
direction goes to zero (whereas the discretization scale in spatial
direction is fixed). Such a limit does indeed work for
discretizations of reparametrization invariant systems and moreover
is suggested by the plots in Figure \ref{fig:3dsmall}.

\begin{figure}[ht]
\begin{minipage}[b]{0.5\linewidth}
\centering
\includegraphics[scale=0.7]{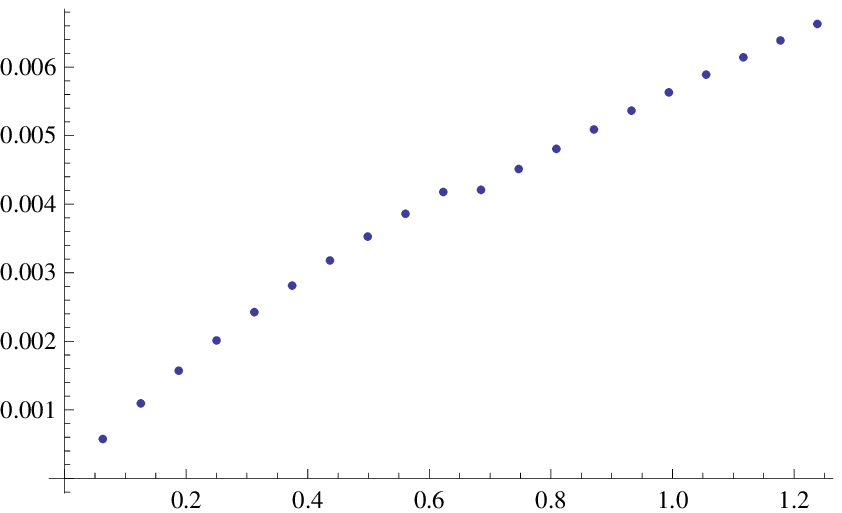}
\caption{The eigenvalue corresponding to translations in time-like
direction as function of the tent pole variable.} \label{3dhamiT}
\end{minipage}
\hspace{0.5cm}
\begin{minipage}[b]{0.5\linewidth}
\centering
\includegraphics[scale=0.7]{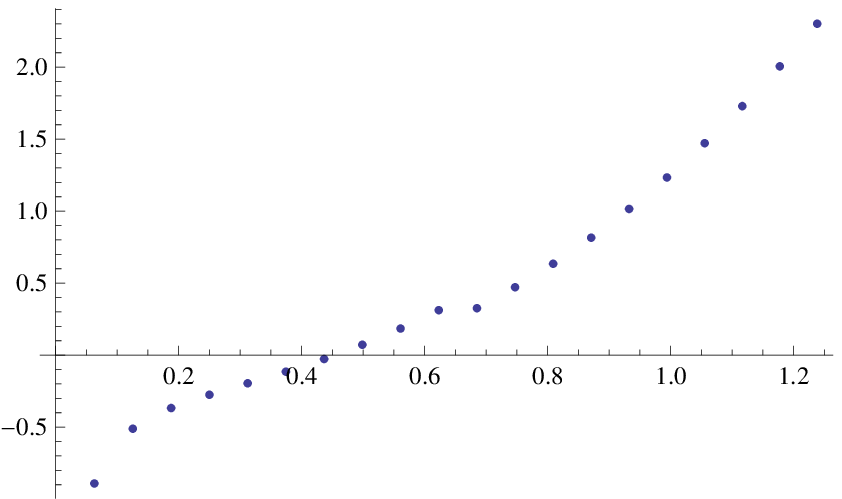}
\caption{  The eigenvalues corresponding to translations in
space-like direction as a function of the tent pole variable. }
\label{3ddiffeoT}
\end{minipage}
\end{figure}

For the system considered here these considerations do not lead to
useful results however. A reason for this can be found by
considering the behavior of the eigenvalues of the Hessian of the
action. That is we consider two consecutive time steps and consider
the Hessian associated to the inner vertex. In Figures
\ref{3dhamiT},\ref{3ddiffeoT} we plot the eigenvalues of this
Hessian, that correspond to translation of the inner vertex in time
like and space-like directions respectively as a function of the
length of the second tent pole. (The cosmological constant is
$\lambda=0.5$, and we again consider a homogeneous solution with
$l^0_i=1,s_{ij}=1$.) The eigenvalue for the time-like direction --
corresponding to the Hamiltonian constraint -- does indeed go to
zero with the length of the tent pole. The eigenvalues corresponding
to the spatial diffeomorphism constraints however start rather to
grow for very small lengths of the tent pole. Hence in this kind of
limit only one of the three gauge symmetries seems to get
restored.\footnote{Note that if we consider a "symmetry reduction"
of the system, in which all the $s_{ij}$ are the same as well as the
$l_i^n$ we reduce the (potential) gauge freedom to translations in
time direction. This is the situation used for the plot in Figure
\ref{fig:3dsmall}. In this case one can indeed find the time
continuous limit and hence a time reparametrization invariant
system, as the gauge freedom for these translations is restored for
infinitesimal time steps. However beginning with the second order in
$\lambda$ the resulting Hamiltonian does not coincide with the
corresponding Hamiltonian constraint obtained from a symmetry
reduction of the alternative dynamic with exact gauge invariance,
see below.} This is in agreement with other arguments \cite{moser}
claiming that for a restoration of diffeomorphism invariance for
Regge calculus one needs to consider triangulations in which the
simplices are "fat" enough, i.e. are not degenerated. But in the
limit we are considering here the tetrahedra become infinitely thin
in time-like directions.

For the dynamics defined by the standard 3d Regge action with a
non--vanishing cosmological constant we do not obtain constraints by
trying to take the limit of a vanishing discretization scale in time
like direction while the spatial discretization scale is kept fixed.
(For actions with an exact gauge invariance such a limit can be
trivially obtained as there arbitrary values for the lapse functions
are allowed and hence also infinitesimal values.) We can also
consider the limit in which both the time-like and space-like
discretization scale is very small. By rescaling the length
variables it is easy to see that this corresponds to taking the
cosmological constant $\lambda$ to be small. Hence to consider this
limit we can perform an expansion of the constraints in $\lambda$.
This expansion also allows us to find the constraints to a certain
order explicitly.

The zeroth order of the constraints is given by the flat dynamics
(\ref{3d2}). To find the first order we will make the ansatz
$t^n={}^0t^n+ \lambda\,{}^1t^n$ in the second equation of
(\ref{3d1}), expand everything to first order and solve for ${}^1
t^n$. We use this result \ba\label{3d5} {}^1t^n=
\left(\frac{\partial \epsilon_t^n}{\partial t^n}\right)^{-1}
\frac{\partial V_{(n,n+1)}}{\partial t^n}_{\,|t^n={}^0t^n} \ea in
the first equation of (\ref{3d1}) and expand again to first order in
$\lambda$: \ba\label{3d6}
p^n_i=-\pi+\theta_{v^ni}(l_i^n,s_{ij})+\lambda\left( \frac{\partial
\psi^{n+}_{i}}{\partial t^n} \left(\frac{\partial
\epsilon_t^n}{\partial t^n}\right)^{-1} \frac{\partial
V_{(n,n+1)}}{\partial t^n}
 - \frac{\partial V_{(n,n+1)}}{\partial l_i^n}   \right)_{\,|t^n={}^0t^n} + O(\lambda^2) \, .\;
\ea Although the explicit expression (\ref{3d6}) as function of the
length variables looks quite lengthy (see appendix \ref{formeln}
detailing the derivatives of simplex volumes and dihedral angles)
the dependence of the first order term on the length variables
$l^{n+1}_j$ at the upper time step drops out. This can be seen by
rewriting the derivative $\frac{\partial \psi^{n+}_{i}}{\partial
t^n}$ in (\ref{3d6}) as \ba\label{3d6b} \frac{\partial
\psi^{n+}_{i}}{\partial t^n}
&=& - \frac{ \partial \theta_{v^ni}^{v^nv^{n+1}ij}}{\partial l_{v^n v^{n+1}}}-\frac{ \partial \theta_{v^ni}^{v^nv^{n+1}ik}}{\partial l_{v^n v^{n+1}}}  \nn\\
&=& - \frac{ \partial \theta_{v^nv^{n+1}}^{v^nv^{n+1}ij}}{\partial l_{v^n i}}-\frac{ \partial \theta_{v^nv^{n+1}}^{v^nv^{n+1}ik}}{\partial l_{v^ni}}  \nn\\
&=& \frac{ \partial \epsilon_t^n}{\partial l_i^n} \ea where we used
that $\frac{\partial \theta^\tau_e}{\partial l_{e'}}=\frac{\partial
\theta^\tau_{e'}}{\partial l_{e}}$ for the dihedral angles
$\theta^\tau_e,\theta^\tau_{e'}$ at the edges $e,e'$ respectively of
a tetrahedron $\tau$ (see appendix \ref{formeln}). In (\ref{3d6b})
the tetrahedron $\tau$ is the one with vertices $v^nv^{n+1}ij$ or
$v^nv^{n+1}ik$ and $e$ the edge between vertices $v^ni$ and $e'$
between $v^nv^{n+1}$.

Now if we denote by ${}^{0}\!t$ the function of the variables
$l_j^n,l_j^{n+1}$ that results by solving the equation
$\epsilon_t^n(t^n,l_j^n,l_j^{n+1})=0$ for $t^n$, we find by taking
the total derivative of this equation with respect to $l_i^n$ that
\ba\label{3d6c} \frac{\partial \,{}^{0}\!t}{\partial
l_i^n}=-\frac{\partial \epsilon_t^n}{\partial l_i^n}
 \left(  \frac{\partial \epsilon_t^n}{\partial t^n}    \right)^{-1}_{\,\,|t^n={}^0t^n}  \q .
\ea
Using (\ref{3d6b}) and (\ref{3d6b}) we have for the first order term in (\ref{3d6})
\ba\label{3d6d}
\left(
\frac{\partial \psi^{n+}_{i}}{\partial t^n}
\left(\frac{\partial \epsilon_t^n}{\partial t^n}\right)^{-1} \frac{\partial V_{(n,n+1)}}{\partial t^n}
 - \frac{\partial V_{(n,n+1)}}{\partial l_i^n}   \right)_{\,|t^n={}^0t^n}
&=&
\left(
-\frac{\partial V_{(n,n+1)}}{\partial t^n}  \frac{\partial \,{}^{0}\!t}{\partial l_i^n}-\frac{\partial V_{(n,n+1)}}{\partial l_i^n}
\right)_{\,|t^n={}^0t^n}  \nn\\
&=&
-\frac{\bd V_{(n,n+1)}({}^{0}\!t(l_j^n,l_j^{n+1}),\,l_j^n,l_j^{n+1})}{\bd l_i^n} \, . \nn\\ \q
\ea

The volume
$V_{(n,n+1)}({}^{0}\!t(l_j^n,l_j^{n+1}),\,l_j^n,l_j^{n+1})$ is the
one of a flat "double pyramid", depicted in Figure \ref{3dtentpic}.
For the kind of orientation of the two tetrahedra as shown in this
figure this volume is just the difference of the volumes of the two
tetrahedra with length $l_j^n,s_{ij}$ and $l_j^{n+1},s_{ij}$
respectively. Therefore the final expression for the momenta to
first order in $\lambda$ is then \ba\label{3d7} p^n_i= -\pi
+\theta_{v^ni}(l^n_j,s_{ij}) +\lambda \frac{\partial
V(l^n_j,s_{ij})}{\partial l^n_i}  +O(\lambda^2) \ea where
$V(l^n_j,s_{ij})$ is the volume of a tetrahedron $\tau$ with edge
lengths $(l_j^n,s_{ij})$. Note that the constraints truncated to
first order are also first class and even Abelian.\footnote{Note
that we did not check the commutator between constraints based at
neighbouring vertices. If the `Cauchy surface' we are considering is
the surface of a tetrahedron, we would get the same first order
constraints for the four three--valent vertices and these would
still be Abelian.} This can be easily seen by realizing that
$S_\tau=\sum_{e \in \tau} l_e(\pi-\theta_e^\tau) -\lambda V^\tau$ is
a generating function for the first order momenta, that is
$p^n_i=-\partial S_\tau/\partial l^n_i+O(\lambda^2)$.

Furthermore the constraints are preserved under time evolution. Note
that the action $S_{(n,n+1)}$ is symmetric under the exchange of the
variables $l^n_i$ and $l^{n+1}_i$. For the first order of the
momenta $p_i^{n+1}$ as defined in the third equation of (\ref{3d1})
we therefore obtain \ba\label{3d8} {}^1\,p^{n+1}_i= \frac{\bd
V_{(n,n+1)}({}^{0}\!t(l_j^n,l_j^{n+1}),\,l_j^n,l_j^{n+1})}{\bd
l_i^{n+1}}  \q , \ea that is again the derivative of the volume
$V_{(n,n+1})=V(l^{n+1}_j,s_{ij})-V(l^n_j,s_{ij})$ of a flat double
pyramid (but this time with a plus sign). We have however to take
the derivative with respect to the length $l^{n+1}_i$, that affects
the larger tetrahedron of this double pyramid. In the end we obtain
the same sign as in (\ref{3d7}) \ba\label{3d9} p^{n+1}_i= -\pi
+\theta_{v^{n+1}i}(l^{n+1}_j,s_{ij}) +\lambda \frac{\partial
V(l^{n+1}_j,s_{ij})}{\partial l^{n+1}_i}  +O(\lambda^2)  \q . \ea

The constraints (\ref{3d8}) coincide to first order with a first
order expansion in $\lambda$ of the constraints describing an exact
discretization of 3d gravity with cosmological constant
\cite{bahrdittrich09b}. The continuum solutions of this theory are
spaces with homogeneous and constant curvature determined by the
cosmological constant. Accordingly instead of a flat tetrahedron
embedded in flat space, we consider a tetrahedron with homogeneously
curved geometry embedded in a space with the same kind of geometry.
In analogy with the constraints (\ref{3d3}) for the flat geometry,
which fix the momenta to agree with the dihedral angles of a flat
tetrahedron, the constraints in this case fix the momenta to agree
with the dihedral angles of a homogeneously curved tetrahedron \be
{}^{exact}p^n_i=-\pi+ {}^\lambda\theta_{v^ni}(l^n_j,s_{ij}) \ee (see
appendix \ref{formeln} for expressions giving the dihedral angles
${}^\lambda\theta_e$ for a tetrahedron in homogeneously curved space
as a function of the edge lengths). An expansion of these
constraints to first order in the curvature or cosmological constant
gives (\ref{3d7}). To see this one needs the identity \be \frac{\bd
\, {}^\lambda\!\theta^\tau_e}{\bd \lambda}= \frac{\partial
\,{}^\lambda\! V}{\partial l_e} \q , \ee where $\,{}^\lambda\! V$ is
the volume of a tetrahedron in homogeneously curved space, that will
be derived in appendix \ref{formeln}.

Starting with second order in $\lambda$ the (pseudo) constraints
(\ref{3d4}) do depend on the length variables $l_i^{n+1}$ at the
upper time step. (This can be more easily checked by considering the
``symmetry reduced'' theory, where all $s_{ij}=s$ and all
$l_i^n=l^n$, see also \cite{bd08}.)

To summarize, we have seen that the canonical equations of motions
for the tent moves reflects the gauge symmetries of the covariant
theory. If the symmetries are exact, we will encounter proper first
class constraints. (See also next section for a general derivation.)
These constraints generate translations of the evolved vertex and
hence mirror the non--uniqueness of the covariant solutions. If the
gauge symmetries are broken, we do not obtain constraints in the
usual sense, as the expressions depend on the lenght variables on
the next time step, or equivalently on lapse and shift. This again
mirrors that the covariant solutions are unique (ignoring discrete
cases of non--uniqueness) and hence lapse and shift are fixed by the
pseudo constraints. However to obtain reasonable (small) values for
lapse and shift, the initial data have to be chosen carefully --
effectively from a ``thickened'' constraint hypersurface of finite
width.

Trying to construct a time continuum limit in order to obtain proper
constraints however fails in the case of 3d Regge calculus with
cosmological constant. The reason is that not all gauge symmetries
are restored in this limit, in which the time discretization scale
goes to zero whereas the spatial discretization scale is fixed.

If we take all edge length to be small, or equivalently consider a
small cosmological constant, we can perform a perturbation in the
cosmological constant $\lambda$. The dynamics truncated to first
order has gauge symmetries and we can obtain constraints. For 3d
Regge calculus with cosmological constant exact constraints
(reflecting exactly the continuum dynamics) exist
\cite{bahrdittrich09b} and are a higher order continuation of the
first order constraints derived here.

We will show in \cite{hoehn} that a similar expansion in curvature
is also possible for 4d Regge calculus. In this case we do not know
the exact (discretized) constraints but if we are able to construct
the first order constraints the question arises if higher order
terms can be derived and whether these are determined uniquely.
These questions will be subjects for further research.

\section{Relation between symmetries of the action and constraints in the canonical framework}\label{relation}

Here we will discuss the relation between gauge symmetries of the action and constraints in more detail.

We argued that gauge symmetries should lead to a non-trivial action
of the gauge group on solutions. As we are considering continuous
groups this should result in non-unique solutions parametrized by
the gauge group parameters. Since solutions are equivalent with
extrema of the action, rather than having one isolated extremum
there is a submanifold of extrema on which the action is constant.
As the first derivatives of the action vanishes by definition we
obtain as a necessary condition for continuous gauge symmetries that
the Hessian of the action should have null directions.

We will consider a triangulation obtained from two consecutive tent
moves. Hence we will have edges with length $l^{n-1}_e$ and
$l^{n+1}_e$ in the `lower' and `upper' boundary of the triangulation
respectively. Here $e$ is an index for the edges from the evolved
vertex $v$ to the adjacent vertices. At the inner vertex $v^n$ will
hinge edges with the length $l^n_e$, $t^{n-1}$ and $t^n$. Denoting
with $S=S_{(n-1,n)}+S_{(n,n+1)}$ the action with boundary terms for
this triangulation the requirement for a null vector
$Y^\alpha,\alpha=e^n,t^{n-1},t^n$ gives the equations
\ba\label{fri1} \sum_e Y^e \frac{\partial^2 S }{\partial l^n_e
\partial l_{e'}^n}
 + Y^{t^{n-1}}  \frac{\partial^2 S }{\partial t^{n-1} \partial l^{e'}}
+ Y^{t^{n}}  \frac{\partial^2 S }{\partial t^{n} \partial l^{e'}}
&=&0   \nn\\
\sum_e Y^e \frac{\partial^2 S }{\partial l^n_e \partial t^{n-1}}
 + Y^{t^{n-1}}  \frac{\partial^2 S }{\partial t^{n-1} \partial  t^{n-1}}
&=&0   \nn\\
\sum_e Y^e \frac{\partial^2 S }{\partial l^n_e \partial t^{n}}
 + Y^{t^{n}}  \frac{\partial^2 S }{\partial t^{n}  \partial t^{n}}
&=&0   \q .
\ea

The momenta conjugated to the tent pole variables have to vanish and
the length of the tent poles do not appear as boundary data. Hence
these variables are not fully dynamical. Therefore we will first
integrate out these variables.

We solve for the lengths of the tent poles $t^{n-1}$ and $t^n$ the
equations of motions \ba\label{fri2}
\frac{\partial S_{(n-1,n)}}{\partial t^{n-1}}&=&0 \nn\\
\frac{\partial S_{(n,n+1)}}{\partial t^{n}}&=&0 \q .
\ea
Using the solutions $T^{n-1}(l_e^{n-1},l^n)$ and $T^{n}(l_e^n,l^{n+1}_e)$ in (\ref{fri2}) and differentiating these identities with respect to $l_e^n$ we obtain
\ba\label{fri3}
\frac{\partial^2 S_{(n-1,n)}}{\partial t^{n-1} \partial l^{n}_e}+ \frac{\partial^2 S_{(n-1,n)}}{\partial t^{n-1} \partial t^{n-1}} \frac{\partial T^{n-1}}{\partial l^{n}_e} &=&0 \nn\\
\frac{\partial^2 S_{(n,n+1)}}{\partial t^{n} \partial  l^{n}_e}+ \frac{\partial^2 S_{(n,n+1)}}{\partial t^{n} \partial t^n} \frac{\partial T^{n}}{\partial l^{n}_e} &=&0  \q .
\ea

The resulting effective action is
\be\label{fri4}
\tilde S(l^{n-1}_e,l^n_e,l^{n+1}_e):=S(l^{n-1}_e,l^n_e,l^{n+1}_e,T^{n-1}(l^{n-1}_e,l^n_e),T^n(l^n_e,l^{n+1}_e) )  \q .
\ee
Using the equations (\ref{fri3}) the Hessian $\tilde H$ of the modified action $\tilde S$ can be written as
\be\label{fri5}
 \frac{\partial^2 \tilde S }{\partial l^n_e \partial l_{e'}^n}
= \frac{\partial^2 S }{\partial l^n_e \partial l_{e'}^n} +
\frac{\partial^2 S}{\partial  l^{n}_e  \partial t^{n-1}}
\frac{\partial T^{n-1}}{\partial l^n_{e'}} + \frac{\partial^2
S}{\partial  l^{n}_e  \partial t^{n}} \frac{\partial T^{n}}{\partial
l^n_{e'}}  \q . \ee Finally with the help of the three equations in
(\ref{fri1}) and again (\ref{fri3}) it follows that null vectors for
the Hessian $H$ of $S$ define also null vectors for the Hessian
$\tilde H$ of $\tilde S$: \be\label{fri6} \sum_e Y^e
\frac{\partial^2 \tilde S }{\partial l^n_e \partial l_{e'}^n} =0 \q
. \ee Because of the second and third equations in (\ref{fri1}) we
can express the components $Y^{t^{n-1}}$ and $Y^{t^n}$ as a
combination of the components $Y^e$. (Here we assume that the second
partial derivatives of the action with respect to $t^{n-1}$ and
$t^n$ do not vanish. This is generically the case for the Regge
action.) For this reason a set of linearly independent null vectors
for the Hessian $H$  will define a set of independent null vectors
for the Hessian $\tilde H$ of the same size.

We will use the action $\tilde S$ to find stationary points
\be\label{fri7}
L^n_e(l^{n-1}_e,l^{n+1}_e)
\ee
as functions of the boundary data $l_e^{n-1},l^{n+1}_e$. Again we use these solutions in the equations of motion
\ba\label{fri8}
\frac{\partial \tilde S}{\partial l^n_e}(l_e^{n-1},L^n_e( l_e^{n-1},l^{n+1}_e  )l^{n+1}) &=& 0
\ea
and differentiate these identities with respect to $l_{e''}^{n+1}$:
\ba\label{fri9}
\sum_{e'} \frac{\partial^2 \tilde S}{\partial l^n_{e} \partial l^n_{e'}} \frac{\partial L^n_{e'}}{\partial l_{e''}^{n+1} }  + \frac{\partial^2 \tilde S_{(n,n+1)}}{\partial l^n_e \partial l_{e''}^{n+1}}=0 \q .
\ea
That is we have
\be\label{fri10}
\tilde H\cdot L=-K
\ee
where $L_{ee'}:=\partial L^n_{e}/ \partial l^{n+1}_{e'}$ and
\be
K_{ee'}:=\frac{\partial^2 \tilde S_{(n,n+1)}}{\partial l^n_e \partial l_{e'}^{n+1}} \q .
\ee\label{fri11}
Therefore null eigenvectors for the Hessian $\tilde H $ are left null eigenvectors for the matrix $K$.

Note that these considerations show that $K$ has at least as many
null vectors as the Hessian $H$. It is not excluded that $K$ has
even more null vectors, however that was not the case in the
examples we investigated.

The matrix $K$ appears as the matrix of derivatives of the momenta
\be\label{fri12} p^n_e:=-\frac{\partial \tilde S_{(n,n+1)}}{\partial
l^n_e} \ee with respect to the length variables $l_{e'}^{n+1}$. If
there exist null vectors we cannot expect to be able to solve the
equations (\ref{fri12}) for the lengths $l^{n+1}_e$ at the next time
step as a function of the momenta and lengths at time $n$ as the
conditions of the implicit function theorem are violated.

If we assume that the rank of the matrix $K$ is constant (and not
maximal) in a neighbourhood of the initial data set
$(l_e^n,l_e^{n+1})$ under consideration then we can conclude that
the Legendre transform from the direct product of configuration
spaces $Q^{n}\times Q^{n+1}$ to the cotangent space $T^*(Q^n)$
defined by \be (l_e^n,l_e^{n+1})\mapsto (l_e^n,p_e^n) \ee is neither
injective nor surjective.

 The image of this map into the phase space is the (primary) constraint hypersurface.
The number of (irreducible) constraints describing this surface is
larger or equal to the number of linearly independent null vectors
for the Hessian of the action in (\ref{fri1}). The constraints are
relations between the momenta and length variables at the same time
step $n$ and will follow from the equations (\ref{fri12}).

This shows that if the action has (local) gauge symmetries, we will
obtain proper constraints on the canonical data. (For general
actions there might also appear constraints without having gauge
symmetries.) Also we cannot solve the relation (\ref{fri12})
uniquely for the length $l^{n+1}_e$ as a function of the data
$l^n_e,p^n_e$ at time step $n$. Hence there will appear arbitrary
parameters (Lagrange multipliers) $\lambda^\alpha$ in the relations
expressing $l^{n+1}$ as a function of $l^n_e,p^n_e$. Whether all
these Lagrange multipliers remain arbitrary or not depends on how
many of the constraints are preserved by time evolution. So we have
the same number of constraints as before following from the
relations \be p^{n+1}_e=\frac{\partial \tilde S_{(n,n+1)}}{\partial
l^{n+1}_e}  \q . \ee

If these constraints coincide with the ones derived from
(\ref{fri12}) then we obtain no further constraints and the Lagrange
multipliers remain arbitrary, as the constraints are automatically
preserved by time evolution. In general it might however happen that
new (secondary) constraints appear or some of the Lagrange
multipliers get fixed, see also \cite{consistent} for an extensive
discussion. For the case considered here --  as the canonical
dynamics is just a reformulation of the covariant Regge equations --
we should obtain the same number of free Lagrange parameters as null
vectors for the Hessian. This was indeed the case for the examples
considered in section \ref{3dcan}.

\section{Discrete reparametrization invariant systems}\label{1d}

We so far discussed examples from Regge calculus where for curved
solutions we did not find an exact form of gauge invariance. One
could argue that the reason for this is that in Regge calculus one
already operates on the space of diffeomorphism invariant
geometries, hence one would not expect any gauge symmetries to
appear.

If we adopt this view it might be however very difficult to
understand the continuum limit. In particular for topological
theories, such as 3d gravity with cosmological constant it is hard
to see how in a refinement limit that introduces more and more
degrees of freedom  (that would all be physical) one can obtain a
continuum theory, in which only a finite and very limited number of
degrees of freedom are actually physical.

Indeed one expects \cite{hartle1,moser,kasner} that in a continuum
limit the (continuum) gauge symmetries are restored. The arguments
are based on the fact that in a refinement limit the ``curvature per
simplex'', that is the deficit angles are getting very small (with
the edge lengths getting small as compared to the average curvature
scale). We have seen that the eigenvalues of the Hessian, that
correspond to the broken or `would--be' symmetries go to zero
(quadratically) with the deficit angles.

The viewpoint that discretizations lead to a breaking of gauge
symmetries that are then restored in the continuum limit is also
strengthened by considering discretizations of parametrized (finite
dimensional) systems. In those cases the reparametrization
invariance is generically broken for the discretized systems. One
could also argue that one is somehow working on the space of gauge
invariant data, there is however always a choice of action for which
reparametrization invariance is fully restored! This shows that it
is important to keep in mind that the question of gauge symmetry or
gauge invariance is one determined by the dynamics -- that is the
action -- of the system.

We will give a short discussion of discretized reparametrization
invariant systems, more material and applications to numerical
integration can be found in \cite{marsden}.

To obtain a continuous action with reparametrization invariance we
can start from a regular Lagrangian $L(q,\dot q)$ where $q$ denote
the configuration variables. To obtain a reparametrization invariant
action we add the time variable $t$ to the configuration variables
and use $s$ as an (auxiliary) evolution parameter instead. It is
then straightforward to verify, that \be\label{action1}
S=\int_{s_i}^{s_f}  L(q, \frac{{q}'}{t'})\, t'\, ds \ee is indeed
invariant under reparametrizations $\tilde s= f(s)$ of the evolution
parameter and the induced change $\tilde t(\tilde s)=t(f^{-1}(\tilde
s)), \tilde q(\tilde s)=q(f^{-1}(\tilde s))$ of the evolution
pathes. Here a prime denotes differentiation with respect to $s$.

One family of discretizations of this action is given by
\be\label{action2}
S_d=\sum_{n=0}^{N-1}  S_{(n,n+1)}
\ee
with
\be\label{action3}
S_{(n,n+1)}:=(t^{n+1}-t^n) \, L((1-\alpha) q^n+\alpha q^{n+1}\,,\,\, \frac{q^{n+1}-q^n}{t^{n+1}-t^{n}})\q .
\ee
The (Euler--Lagrange) equations of motion are then given by second order difference equations which can be obtained from
\ba\label{action4}
0&=&\frac{ \partial S_{(n-1,n)}}{\partial t^n} + \frac{ \partial S_{(n,n+1)}}{\partial t^n}   \nn\\
0&=&\frac{ \partial S_{(n-1,n)}}{\partial q^n} + \frac{ \partial S_{(n,n+1)}}{\partial q^n}      \q  .
\ea

As an example consider the discretized reparametrization invariant harmonic oscillator with a discrete action for one time step given by
\be\label{action5}
S_{(n,n+1)}=\left( \frac{1}{2} \frac{ (q^{n+1}-q^n)^2}{(t^{n+1}-t^n)^2} - \frac{1}{2} w\left(\tfrac{1}{2} q^{n}+\tfrac{1}{2}q^{n+1}\right)^2\right)(t^{n+1}-t^n)  \q
\ee
where $w$ parametrizes the strength of the potential term.

Assume we consider the smallest boundary value problem with
prescribed data at time steps $(n-1)$ and $(n+1)$. A necessary
condition for the system to be reparametrization invariant is that
the Hessian of the action with respect to the variables at the time
step $n$ has a null vector if evaluated on solutions. Now basically
the same discussion as in section \ref{relation} will show that if
this Hessian has a null vector, this is also the case for the matrix
\be\label{action6} K:=\left( \begin{array}{cc}
\frac{\partial^2 S_{(n,n+1)}}{\partial q^n \partial q^{n+1}} & \frac{\partial^2 S_{(n+1)}}{\partial q^n \partial t^{n+1}}  \\
\frac{\partial^2 S_{(n,n+1)}}{\partial t^n \partial q^{n+1} } & \frac{\partial^2 S_{(n,n+1)}}{\partial t^n \partial t^{n+1}}
\end{array} \right)  \q.
\ee

(To consider this matrix has the advantage that we do not need to
determine the solution.) An explicit calculation gives for the
determinant of $K$ \be\label{action7} \det(K)=\frac{1}{4} w
\frac{(q^{n+1}-q^n)^2}{(t^{n+1}-t^n)^2}+ \frac{1}{4} w^2
(\tfrac{1}{2}q^{n+1}+\tfrac{1}{2}q^n)  \q , \ee that is a
non--vanishing result. Hence the discrete system does not display
(local) gauge symmetries. Note that for a free particle, $w=0$,
reparametrization invariance persists for the discretized theory. As
we will see later on the reason is that the dynamic for a free
particle defined by the discretized theory coincides with the
dynamic of the continuum theory.

This raises the question whether for instance similarly to the case
of 3d gravity with cosmological constant discussed in section
\ref{3dcan}, the gauge symmetry and therefore constraints persist
for a first order truncation in $w$. It will turn out that this is
not the case. On the other hand one can define the time continuum
limit without any problems and recover reparametrization invariance.

To define time evolution in a canonical formalism we again use the discrete action for one time slice as a generating function:
\ba\label{action8}
p_q^n &=& - \frac{ \partial S_{(n,n+1)}}{\partial q^n} = \frac{(q^{n+1}-q^n)}{(t^{n+1}-t^n)}+\tfrac{1}{2}w(\tfrac{1}{2} q^{n+1}+\tfrac{1}{2}q^n)(t^{n+1}-t^n)\nn\\
p_t^n&=& -\frac{ \partial S_{(n,n+1)}}{\partial t^n} =-\frac{1}{2}\frac{(q^{n+1}-q^n)^2}{(t^{n+1}-t^n)^2}-\tfrac{1}{2}w(\tfrac{1}{2} q^{n+1}+\tfrac{1}{2}q^n)^2   \nn\\
p_q^{n+1} &=& \q \frac{ \partial S_{(n,n+1)}}{\partial q^{n+1}} = \frac{(q^{n+1}-q^n)}{(t^{n+1}-t^n)}-\tfrac{1}{2}w(\tfrac{1}{2} q^{n+1}+\tfrac{1}{2}q^n)(t^{n+1}-t^n) \nn\\
p_t^{n+1}&=& \q \frac{ \partial S_{(n,n+1)}}{\partial t^{n+1}}=-\frac{1}{2}\frac{(q^{n+1}-q^n)^2}{(t^{n+1}-t^n)^2}-\tfrac{1}{2}w(\tfrac{1}{2} q^{n+1}+\tfrac{1}{2}q^n)^2  \q .
\ea

For the free particle $w=0$ the first two equations in
(\ref{action8}) cannot be solved uniquely for $(q^{n+1},t^{n+1})$
and one is left with a one-parameter ambiguity. Accordingly there is
one constraint \be\label{action9} C_{w=0}=p_t^n +\tfrac{1}{2}
(p_q^n)^2=0 \ee between the canonical data at any time step $n$. For
$w >0$ there are at most  discrete ambiguities in the solutions
$(q^{n+1},t^{n+1})$ as functions of the canonical data at time step
$n$. To check whether we obtain a constraint in a first order
truncation in $w$ we expand \ba q^{n+1}=\,{}^0q^{n+1}+ w \,{}^1
q^{n+1}  \q , \q\q t^{n+1}={}^0t^{n+1}+ w \,{}^1 t^{n+1} \ea where
${}^0t^{n+1}=t^n+ (\,{}^0p_q^n)^{-1}(\,{}^0q^{n+1}-q^n)$ is the
solution for the $w=0$ dynamics. We use these expansions in the
defintion of the momenta $p_q^n,p_t^n$ in (\ref{action8}) and
neglect any higher than first order terms in $w$. From the resulting
equations \ba\label{action10}
p_q^n &=&\,{}^0 p_q^n - w\frac{(\,{}^{1}q^{n+1}-q^n)^2(\,{}^{1}q^{n+1}+q^n)  + 4 (\,{}^0p_q^n)^2(\,{}^0p_q^n \,\,{}^1t^n-\,\,{}^1q^n   )  }{ 4  \,\,{}^0p_q^n (\,{}^{1}q^{n+1}-q^n)    } \nn\\
p_t^n&=&-\tfrac{1}{2} ( \,{}^0 p_q^n  )^2+w
\frac{(\,{}^{1}q^{n+1}-q^n)(\,{}^{1}q^{n+1}+q^n)^2  + 8
(\,{}^0p_q^n)^2(\,{}^0p_q^n \,\,{}^1t^n-\,\,{}^1q^n   )  }{ 8
(\,{}^{1}q^{n+1}-q^n)    } \ea we do not obtain a constraint. Rather
these can be solved for instance for the variables ${}^1 q^{n+1}$
and ${}^0 q^{n+1}$ (with ${}^1t^{n+1}$ remaining undetermined). That
is these (first order) equations fix the values for the zeroth order
variables ${}^0q^{n+1},\, {}^0 t^{n+1}$ which were only determined
up to a one-parameter reparametrization invariance by the zeroth
order equations. This behaviour differs from the first order
dynamics in the cosmological constant for 3d Regge calculus found in
section \ref{3dcan}.

On the other hand it is possible to find a continuous time limit, in
which reparametrization invariance is restored. (This was not
possible for 3d Regge calculus with cosmological constant.) Again
expanding \ba q^{n+1}=q^n+ \varepsilon\, \,{}^1 q^{n+1}  \q , \q\q
t^{n+1}=t^{n}+ \varepsilon \,\,{}^1 t^{n+1} \ea we will find that up
to second order terms the relation \be \label{action11}
C_\varepsilon=p^n_t+\tfrac{1}{2}(p_q^n)^2+\tfrac{1}{2}w (q^n)^2
+O(\varepsilon^2)=0 \ee between the canonical data. Indeed
(\ref{action11}) coincides with the continuum constraint for the
parametrized harmonic oscillator.

Finally we want to point out that for such mechanical systems, in
which we have only a one--dimensional reparametrization invariance
in time direction, it is straightforward to define a discrete
dynamics which preserves reparametrization invariance.
 The idea is that the discrete system
should exactly reproduce the dynamics of the continuous system. To
achieve such a dynamics one can start with the continuum action and
solve for all variables $t(s),q(s)$ except for those at some
discrete subsets of the evolution parameter $s_n$. Reinserting the
solutions in the action will result in a sum of Hamilton--Jacobi
functions
\begin{eqnarray} \label{actionb1}
S_{exact}\;=\;\sum_{n=0}^{N-1}
S_{HJ}^{(s^n,s^{n+1})}(t^n,q^n,t^{n+1},q^{n+1})
\;=\;\sum_{n=0}^{N-1}\int_{s_n}^{s_{n+1}}ds\; {\cal L} (t(s),q(s))\q .
\end{eqnarray}
where ${\cal L}$ is the Lagrangian of the reparametrization invariant continuous system, and $t(s),q(s)$ are solutions with boundary data $(t^n,q^n,t^{n+1},q^{n+1})$.

If we define the momenta for this discretized dynamics according to (\ref{action8}) we actually obtain the same definition as for the continuums dynamics
\ba
p^n_q&=& -\frac{\partial S_{HJ}^{(s^n,s^{n+1})}(t^n,q^n,t^{n+1},q^{n+1})}{\partial q^n} \nn\\
p^n_t&=& -\frac{\partial
S_{HJ}^{(s^n,s^{n+1})}(t^n,q^n,t^{n+1},q^{n+1})}{\partial t^n} \q .
\ea Hence the same constraints between the momenta and configuration
variables as for the continuum dynamics have to hold, which also
shows that reparametrization invariance is preserved.

\section{Discussion}\label{discussion}

We have shown that we do not have exact gauge symmetries for
non--flat vertices in discretized 4d gravity as defined by the Regge
action. In a canonical formulation for Regge calculus this will lead
to pseudo constraints\footnote{Note that for instance for
four-valent tent moves or Cauchy surfaces formed by the boundary of
a 4--simplex one will obtain proper constraints restricting the
physical degrees of freedom to zero, as the dynamics supported by
these structures is just given by 4d flat space, see
\cite{dittrichryan}.} involving data at different time steps instead
of proper constraints involving data at the same time step.

This fact should be considered in attempts to connect covariant and
canonical quantum gravity models, for instance loop quantum gravity,
based on proper constraints, and spin foam models, that can be seen
as quantizations of Regge calculus, which rather lead to pseudo
constraints. Indeed the methods discussed in this paper allow to
introduce a canonical formalism that completely matches the dynamics
of the covariant system and moreover  using tent moves makes a local
analysis possible. These techniques seem therefore be ideally suited
to be applied to other discretized actions, as the discretized
Plebanski action \cite{plebanski} used in spin foam models, or the area-angle
action for Regge calculus introduced in \cite{dittrichspeziale}.
This could help to derive a closer connection between spin foam
models and loop quantum gravity.

In the case that one starts with an action with broken symmetries
and obtains pseudo constraints, there might nevertheless exist
certain limiting cases in which these turn into proper constraints.
This was the case for the first order dynamics in the cosmological
constant for 3d Regge calculus. An analysis for 4d Regge calculus
for such limiting cases will appear in \cite{hoehn}. Starting from
these proper constraints it might be possible to extend the
constraints, such that in the end one obtains a system with an
alternative dynamics with exact gauge symmetries.

The example of reparametrization invariant systems shows that the
breaking of the symmetries in discrete theories is due to deviations
of the discrete dynamics from the continuum one. We discussed
examples, where different discretizations of the action lead to a
symmetry breaking and symmetry preserving dynamics. Hence one should
be careful in deriving general conclusions for quantum gravity
models from the fact that the symmetries are broken for the Regge
action. Alternative actions might display exact gauge invariance, see also \cite{nielsen} for attempts to obtain such actions.

For deparametrization invariant systems we have seen that such an
alternative action can be always defined by starting from the
continuum action and integrating out infinitely many degrees of
freedom, that is in other words by applying a renormalization group
transformation. We will explore this idea in more detail in
\cite{bahrdittrich09b}. Indeed in lattice field theories the so-called
perfect actions, which reflect exactly the continuum dynamics, can
be constructed as fixed points of the renormalization group flow
\cite{perfect}. These lead typically to non--local couplings (i.e.
couplings extending over non--adjacent lattice sites). It would be
therefore interesting to extend the analysis of this paper to
non--local actions.

Note that another property for the exact discretized action
(\ref{actionb1}) is the independence under refinements (for the
partition function or dynamical observables). Indeed as the dynamics
already reflects the continuum dynamics, there is nothing gained by
refining the lattice. Similarly one would expect that a perfect
action for discretized gravity - should it exist - would lead to
triangulation independent partition functions. This is well known
for topological field theories (such as 3d gravity), where however
the actions are still local. For theories with propagating degrees
of freedom non--local couplings might arise (even at the classical
level).

An alternative way to obtain triangulation independence for instance
for spin foam models, is to sum over triangulations, which would
also include infinitely refined triangulations. This can however be
only defined at a formal level, for instance by invoking group field
theory methods \cite{daniele}. In some sense the construction of a
perfect action can be seen as taking the effects of a sum over
triangulation into account and might in this way circumvent
divergencies that appear in a sum over triangulations. An
investigation into the properties of such actions seems therefore
promising to us.

\appendix

\section{An example for physically different solutions with the same boundary conditions}\label{appamb}

In this section we will construct an example of two physically
different Regge solutions for the same boundary data. The
construction uses two tent moves. In the end we will obtain two
solutions, one flat and one with curvature. As the final
triangulation has an inner vertex we will check the Hessian of the
action evaluated on these solutions and find null eigenvalues for
the flat solution and only non--vanishing eigenvalues for the curved
solution.

\subsection{One tent move}

We consider the following four-dimensional triangulation: start with
a regular tetrahedron where all edges have unit length. Apply a 1-4
move, i.e. add another vertex $A$, which is connected to each of the
four former vertices $V_1,\ldots,V_4$. The length of each of the
edges from $A$ to $V_i$ is $a>0$. This three-dimensional
triangulation serves as a starting point for a tent move applied at
$A$. The resulting four-dimensional triangulation has yet another
vertex $B$, which is connected with each of the $V_i$, the length of
each of the connecting edges being $b$. Denote the length of the
tent pole, i.e. the edge between $A$ and $B$ by $t$.
\begin{figure}[hbt!]\label{fig:OneFourMove}
\begin{center}
    \psfrag{1}{$V_1$}
    \psfrag{2}{$V_2$}
    \psfrag{3}{$V_3$}
    \psfrag{4}{$V_4$}
    \psfrag{l1}{$1$}
    \psfrag{l2}{$a$}
    \psfrag{A}{$A$}
    \includegraphics[scale=0.6]{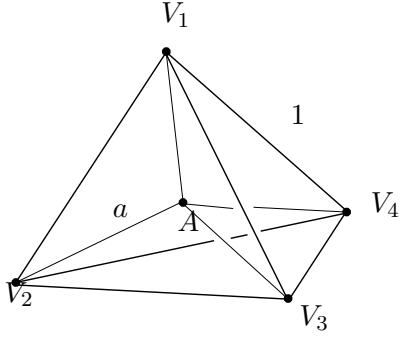}
    \end{center}
    \caption{\small The initial triangulation. The lengths of the edges connecting any $V_i$ to any other $V_j$ is $1$, the length of the edges connecting $A$ to any of the $V_i$ is $a$.}
\end{figure}

The tent pole is the only inner edge of this triangulation, and its
Regge equation reads

\begin{eqnarray}
\sum_{i=1}^4\frac{\partial A_{ABV_i}}{\partial
t}\,\epsilon_{ABV_i}\;=\;0,
\end{eqnarray}

\noindent where $A_{ABV_i}$ is the area of the triangle between the
vertices $A$, $B$ and $V_i$, and $\epsilon_{ABV_i}$ is the deficit
angle at the corresponding triangle. Due to our symmetric setting,
neither of these depend on $i$, and is solved if either of the
following equations holds:

\begin{eqnarray}\label{Gl:ReggeForTentPole1}
\frac{\partial A_{ABV_i}}{\partial t}\;&=&\;\frac{t(a^2+b^2-t^2)}{2
\sqrt{-t^4-(a^2-b^2)^2+2t^2(a^2+b^2)}
}\;=\;0\\[5pt]\label{Gl:ReggeForTentPole2}
\epsilon_{ABV_i}\;&=&\;2\pi\,-\,3\arccos\left[\frac{-(a^2-b^2)^2+2(a^2+b^2-1)t^2-t^4}{2(a^4+t^2+(b^2-t^2)^2-2a^2(b^2+t^2))}\right]\;=\;0.
\end{eqnarray}

\noindent The equation (\ref{Gl:ReggeForTentPole1}) has
$t=t_\perp:=\sqrt{a^2+b^2}$ as a solution. The only two positive
solutions for the second equation are
\begin{eqnarray}\label{Gl:FlatTentPole}
t\;=\;t_\pm\;:=\;\frac{1}{2}\sqrt{4a^2+4b^2-3\pm\sqrt{8a^2-3}\sqrt{8b^2-3}}
\end{eqnarray}

\noindent The two latter solutions $t_\pm$ can readily be identified
as the two flat solutions pointing in `forward' and `backward' direction. (For $b>a$ the solution $t_-$ is forward pointing, whereas $t_+$ is backward pointing.)\\[5pt]

\subsection{Two tent moves}

We now again start with the three-dimensional triangulation shown in
figure \ref{fig:OneFourMove}, but successively apply two tent moves
to the vertex $A$. The resulting triangulation has an inner vertex,
$B$, connected to the vertices $V_i$, the length denoted by $b$. The
final spatial triangulation contains a vertex $C$, the lengths of
the edges connecting any of the $V_i$ to $C$ being of length $c$.
For simplicity, we choose $c=a$ for the moment. Denote the lengths
of the tent pole connecting $A$ and $B$, and $B$ and $C$ by $s_1$
and $s_2$ respectively. Due to symmetry, the Regge equations for
$s_1,s_2$ can immediately be solved to be either of
\begin{eqnarray}\label{Gl:SolutionsForTentPoles}
s_{1,2}\;=\;t_\perp,\,t_\pm
\end{eqnarray}

\noindent with $t_\perp:=\sqrt{a^2+b^2}$ and $t_\pm$ given by
(\ref{Gl:FlatTentPole}). Note that there is still one more equation
to solve, which is the equation for $b$, the bulk edge connecting
$V_i$ and $B$. The Regge equation for $b$ reads
\begin{eqnarray}\label{Gl:ReggeForB}
-\frac{\partial S}{\partial b}\;=\;\frac{\partial A_{ABV_i}}{\partial
b}\epsilon_{ABV_i}\;+\;\frac{\partial A_{CBV_i}}{\partial
b}\epsilon_{CBV_i}\;+\;3\frac{\partial A_{BV_iV_j}}{\partial
b}\epsilon_{BV_iV_j}\;=\;0
\end{eqnarray}

\noindent The equation (\ref{Gl:ReggeForB}) will have a different
form for each of the six choices (\ref{Gl:SolutionsForTentPoles})
for the lengths $s_1,s_2$, and result in different equations for
$b$.

We first consider the cases where $s_1,s_2=t_\pm$. As we have
already seen, this results in the deficit angles
$\epsilon_{ABV_i}=\epsilon_{CBV_i}$ to vanish. Since for any $i\neq
j$ one has
\begin{eqnarray}
\frac{\partial A_{BV_iV_j}}{\partial
b}\;=\;\frac{b}{2\sqrt{4b^2-1}},
\end{eqnarray}

\noindent the only way (nonzero $b$)  for (\ref{Gl:ReggeForB}) to be
satisfied is that also $\epsilon_{BV_iV_j}=0$, so the overall
triangulation is flat.

It is now not hard to show -- in case that either $s_1=t_+,s_2=t_-$
or $s_1=t_-,s_2=t_+$ -- that $\epsilon_{BV_iV_j}$ is identically
zero for $b$ lying in some finite interval (which, after some
investigation, can be found to be the range of values such that all
triangles in the triangulation satisfy the triangle inequalities).
This gives a one parameter family of flat solutions for the two
cases. In the case that either $s_1=s_2=t_+$ or $s_1=s_2=t_-$, the
only positive solution is
\begin{eqnarray}
b\;=\;\sqrt{\frac{3}{8}},
\end{eqnarray}
\noindent which is the point where $t_+=t_-$, so this is just a
special case of the flat
solution.

Next we consider $s_1=t_+$ and $s_2=t_\perp$. Again, one finds that
the deficit angle $\epsilon_{BV_iV_j}$ has to vanish, resulting in a
completely flat triangulation. The only positive solution is
\begin{eqnarray}
b\;=\;\frac{\sqrt 3 a}{\sqrt{8a^2-3}},
\end{eqnarray}

\noindent being the solution of $t_\perp=t_-$, hence again a special
case of $s_1=t_+$ and $s_2=t_-$. The case $s_1=t_-$ and
$s_2=t_\perp$ works analogously.

There is one case left, which is $s_1=s_2=t_\perp =\sqrt{a^2+b^2}$.
Since neither of the deficit angles has to vanish now, the Regge
equation (\ref{Gl:ReggeForB}) involves multiple terms, therefore is
genuinely a transcendental equation, given by
\begin{eqnarray}\nonumber
\textstyle
a\left(2\pi-3\arccos\frac{-b^2+a^2(2b^2-1)}{b^2+a^2(4b^2-1)}\right)\,+\,\frac{3b}{2\sqrt{4b^2-1}}\left(2\pi-4\arccos\frac{b^2}{\sqrt{(3b^2-1)(-b^2+a^2(4b^2-1))}}\right)\;=\;0\, .\\ \label{Gl:ReggeForBPerpindicular}
\end{eqnarray}

\begin{figure}[hbt!]
\begin{center}
    \includegraphics[scale=0.6]{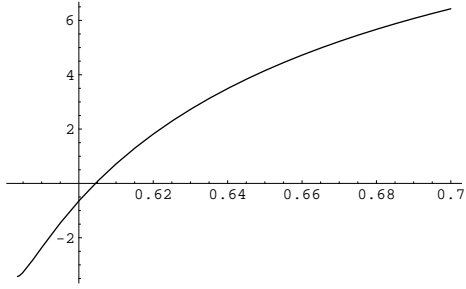}
    \end{center}
    \caption{\small The derivative  $-\partial S/\partial b$ as a function of $b$.}\label{fig:CurvedSolution}
\end{figure}

\noindent It still can be solved numerically. For $a=4$, one can
plot the derivative $-\partial S/\partial b$ w.r.t $b$, see Figure \ref{fig:CurvedSolution}.

\noindent The zero can be evaluated numerically to be $b\approx
0.604458$. The deficit angles can be computed numerically for this
case as well, and are all found to be unequal to zero:
\begin{eqnarray}
\epsilon_{ABV_i}\;=\;\epsilon_{CBV_i}\;\approx\;-0.615334 \, , \q
\epsilon_{BV_iV_j}\;&\approx&\;1.84412 \, .
\end{eqnarray}

\noindent The four-dimensional volume of the triangulation is
computed to be
\begin{eqnarray}
V\;=\;8V_{ABV_iV_jV_k}\;\approx\;0.180461
\end{eqnarray}

\noindent Similarly, the three-dimensional volumes of the tetrahedra
can be computed and found to be positive, as well as all areas of
triangles. Therefore, all generalized triangle inequalities are
satisfied and the solution defines a geometrical
triangulation.

Note that - with the same boundary conditions $a=c=4$, there could
always be chosen the flat solution $s_1=t_+,s_2=t_-$, e.g. with
$b=\sqrt{3/8}$, for definiteness. The corresponding volume can be
computed to be
\begin{eqnarray}
V\;=\;8V_{ABV_iV_jV_k}\;\approx\;0.232924 \, .
\end{eqnarray}

\noindent which is different from the curved case.\\

This gives an explicit example where the the specification of
boundary values\footnote{In this case we chose $a=c=4$ for
simplicity, however the same type of curved solution exist  also for
$a\neq c$.}, and a fixed triangulation (containing an inner vertex)
does not result in a unique solution. Apart from the flat solution,
there is a curved one as well. Also, this curved solution is
isolated, i.e. there is no continuous symmetry of the solution,
which we now confirm by computing the Hessian of the action at the
solution.

In our triangulation there are six edges meeting at the vertex $B$.
\begin{figure}[hbt!]\label{fig:Hessian}
\begin{center}
    \psfrag{1}{$V_1$}
    \psfrag{2}{$V_2$}
    \psfrag{3}{$V_3$}
    \psfrag{4}{$V_4$}
    \psfrag{b1}{$b_1$}
    \psfrag{b2}{$b_2$}
    \psfrag{b3}{$b_3$}
    \psfrag{b4}{$b_4$}
    \psfrag{s1}{$s_1$}
    \psfrag{s2}{$s_2$}
    \psfrag{A}{$A$}
    \psfrag{B}{$B$}
    \psfrag{C}{$C$}
    \includegraphics[scale=0.6]{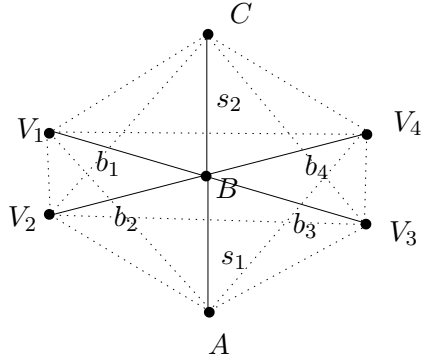}
    \end{center}
    \caption{\small The 4d triangulation with inner vertex $B$. }
\end{figure}
\noindent Therefore, the Hessian is a $6\times 6$ matrix the entries
being the second derivatives of the action. Since our solution is
highly symmetric, with $b_1=\ldots, b_4=b$ and $s_1=s_2$, 
%
%
%
%
%
%
%
%
\noindent some of the eigenvalues coincide
\begin{eqnarray*}
\lambda_1\;&\approx&\;-137.752,\quad
\lambda_2\;\approx\;-8.32992,\quad\lambda_3\;\approx\;-8.32393\\[5pt]
\lambda_4\;&=&\;\lambda_5\;=\;\lambda_6\;\approx\;0.339397 \; .
\end{eqnarray*}
\noindent One can see that the Hessian possesses no zero eigenvalue,
showing that there is no infinitesimal symmetry of this solution. It
is instructive to compare this solution with the flat, homogenous
solution, which is given by
\begin{eqnarray}
b=\sqrt{\frac{3}{8}},\,s_1\;=\;s_2\;=\;\sqrt{a^2-\frac{3}{8}} \q .
\end{eqnarray}
\noindent 
In this case there are four null eigenvectors that can be associated to translations of the inner vertex in the four directions.

\section{Geometric relations in simplices}\label{formeln}

Consider a $D$--dimensional simplex in $D$--dimensional Euclidean
space or in $D$--dimensional manifold of constant sectional
curvature $\kappa\neq 0$ consisting of $D+1$ vertices $v_1,\ldots,
v_{D+1}$. Denote this simplex by $(123\ldots D+1)$. Any subsimplex
is determined by the subset $v_{i_1},\ldots,v_{i_n}$ of the vertices
which span this subsimplex, and will therefore be denoted as
$(i_1i_2,\ldots,i_n)$.\footnote{The order of the $i_k$ determines
also an orientation of the subsimplex, which however plays a minor
r\^ole in this article.} Similarly, the notion $(\hat i_1\hat
i_2\ldots,\hat i_n)$ is denoting the subsimplex which consists of
all vertices other than $i_1,\ldots,i_n$. The subsimplices in curved
space are defined to be the hypersurfaces with zero extrinsic
curvature as embeddings in the geometry of the higher dimensional
simplex. These are in fact also simplices of curvature $\kappa$. An
edge $(ij)$ is then just given by the geodesic connecting $v_i$ and
$v_j$.\footnote{Note that if $\kappa>0$ then there are at least two
such geodesics. In fact, there are several different curved
simplices which have the same vertices. For definiteness we will always choose
the shorter of the two geodesics, and not consider cases in which
there are infinitely many such geodesics, which are degenerate
anyway.}

\begin{figure}[ht]
\begin{minipage}[b]{0.5\linewidth}
\centering
    \psfrag{12}{$(12)$}
    \psfrag{23}{$(23)$}
    \psfrag{34}{$(34)$}
    \psfrag{t}{$\theta_{34}$}
    \psfrag{14}{$(14)$}
    \psfrag{1}{$v_1$}
    \psfrag{2}{$v_2$}
    \psfrag{3}{$v_3$}
    \psfrag{4}{$v_4$}
    \includegraphics[scale=0.5]{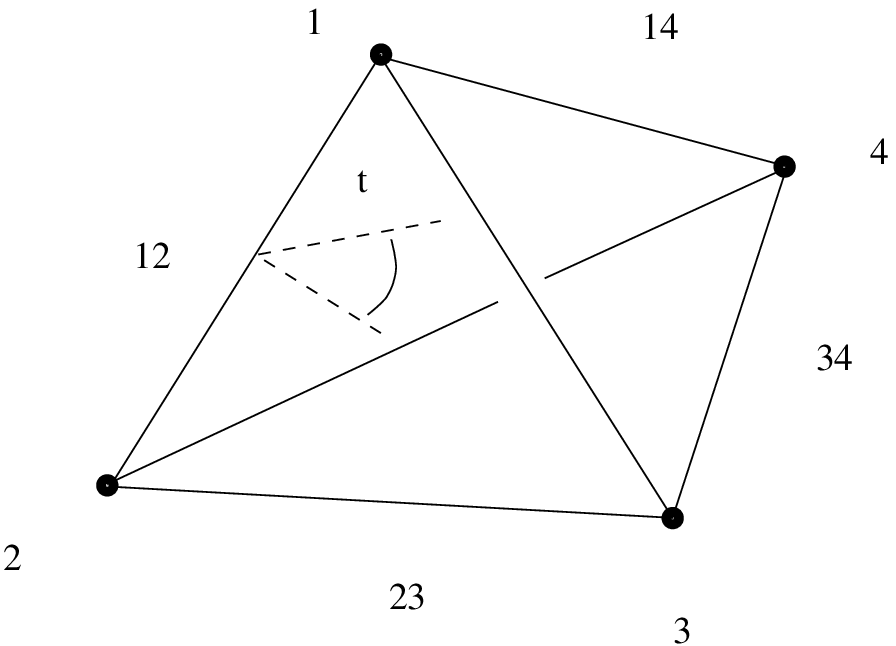}
\caption{\small The interior dihedral angle in the tetrahedron
$(1234)$ at the edge $(\hat3\hat4)=(12)$.} \label{fig:InteriorAngle}
\end{minipage}
\hspace{0.5cm}
\begin{minipage}[b]{0.5\linewidth}
\centering
 \psfrag{12}{$(12)$}
    \psfrag{1}{$v_1$}
    \psfrag{2}{$v_2$}
    \psfrag{3}{$v_3$}
    \includegraphics[scale=0.5]{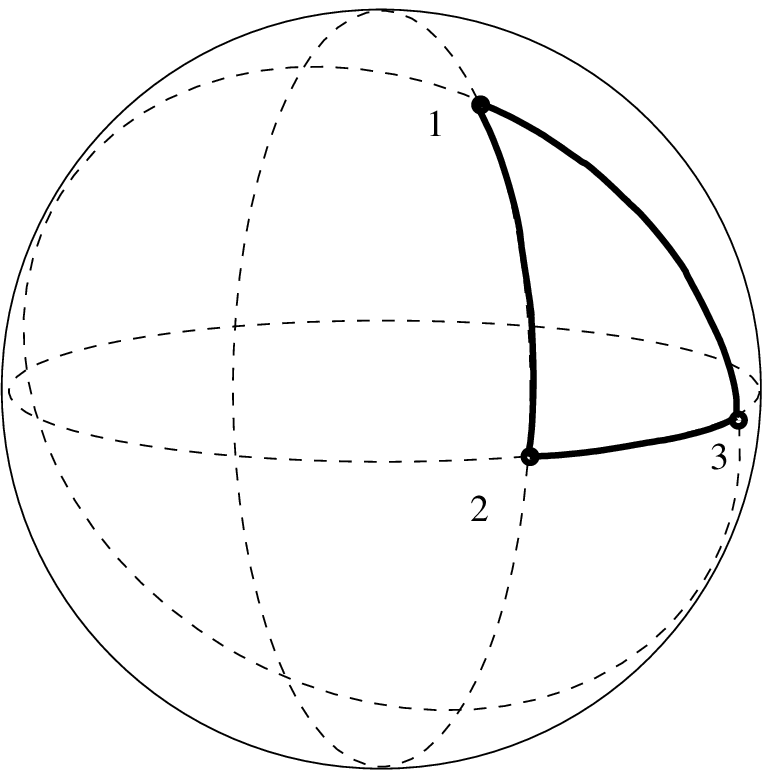}
    \caption{\small A spherical triangle in $S^2$ is a simplex of constant curvature $1$. Note that the edges are simplices in $S^1$, which are the great circles, having zero extrinsic curvature in
    $S^2$.}\label{fig:SphericalTriangle}
\end{minipage}
\end{figure}

\noindent Denote the geodesic lengths of the edges $(ij)$ by
$l_{ij}$. Then the $(D+1)\times(D+1)$ matrix $G$ with entries
\begin{eqnarray}
G_{ij}\;=\;c_\kappa(l_{ij})
\end{eqnarray}

\noindent where the function $c_\kappa(x)$ is defined by
\begin{eqnarray*} \label{ckappa}
c_\kappa(x)\;:=\;\left\{\begin{array}{ll}
\cos\big(\sqrt\kappa x\big)&\quad \kappa>0\\[5pt]
\cosh\big(\sqrt{-\kappa} x\big)&\quad
\kappa<0\end{array}\right.
\end{eqnarray*}

\noindent is called the \emph{Gram matrix} of the simplex. For $\kappa=0$ we define
\be
G_{ij}=-\tfrac{1}{2} \sum_{k,l}l^2_{kl} (\delta _{ik}-\tfrac{1}{D+1})(\delta_{jl}-\tfrac{1}{D+1}) .
\ee
to be the affine metric of the simplex, see for instance \cite{bdlfss}.
For $\kappa\neq 0$ we denote by $c_{ij}$ the $ij$-th cofactor of
$G$, i.e. the determinant of the matrix obtained by removing the
$i$-th row and $j$-th column of $G$. For the flat case $\kappa=0$,
we take $c_{ij}$ to be the affine inverse metric, that is $\sum_k
G_{ik}c_{kj}=\delta_{ij}-\frac{1}{D+1}$.\\ Then the interior
dihedral angle $\theta_{[ij]}$ opposite of the edge $(ij)$ is given
by \cite{kock,bdlfss}
\begin{eqnarray}\label{dihedral}
\cos\theta_{[ij]}\;=\;-\frac{c_{ij}}{\sqrt{c_{ii}}\sqrt{c_{jj}}}.
\end{eqnarray}

\noindent Denote the volume of the
subsimplex spanned by all vertices except $v_i$ and $v_j$ by
$V_{(\hat i\hat j)}$. For variations $\delta$ of the geometry of a simplex the \emph{Schl\"afli-identity} \cite{BookMilnor}
\begin{eqnarray}\label{Gl:SchlaefliIdentityForCurvedSimplices}
\sum_{i<j}V_{(\hat i\hat j)} \delta \theta_{[ij]}\;=\;(D-1)\,\kappa\,\delta V_{(12\ldots
D+1)}
\end{eqnarray}
 holds.

With the help of this identity we can derive some formulas needed in
section \ref{3dcan}. First of all if for a given  tetrahedron we
define the function \be S_\tau=\sum_{i<j}V_{(\hat i\hat j)}
\theta_{[ij]} -2 \,\kappa\, V_{(12\ldots D+1)}  \q \ee we can use it
as a generating function for the dihedral angles \be
\theta_{[ij]}=\frac{ \partial S_\tau} {\partial V_{(\hat i \hat j)}}
\q \ee where $V_{(\hat i \hat j)}$ is the length of the edge between
the two other vertices $k,l\neq i,j$ of the tetrahedron. Since
partial derivatives commute  we can conclude \be\label{commu1}
\frac{ \partial \theta_{[ij]}}{\partial V_{(\hat k\hat l)} }= \frac{
\partial \theta_{[kl]}}{\partial V_{(\hat i  \hat j)} } \ee which is
used in section \ref{3dcan}, for instance to see that the
constraints derived there are Abelian.

To compute the derivative of the dihedral angles with respect to the
sectional curvature $\kappa$, also needed in section \ref{3dcan}, we
note from the definition (\ref{ckappa},\ref{dihedral}) of the
dihedral angles that \ba
\frac{\bd \theta_{[kl] }}{\bd \kappa} &=& \sum_{i>j} \frac{ \partial \theta_{[kl]}}{\partial V_{(\hat i\hat j)}} \frac{1}{2} \kappa^{-1} \, V_{(\hat i \hat j)} \nn\\
 &=& \sum_{i>j} \frac{ \partial \theta_{[ij]}}{\partial V_{(\hat k\hat l)}} \frac{1}{2} \kappa^{-1} \, V_{(\hat i \hat j)} \nn\\
&=&  \frac{\partial V_{(1234)}}{\partial V_{(\hat k \hat l)}}
\ea
where in the second line we use (\ref{commu1}) and in the third line the Schl\"alfi identity (\ref{Gl:SchlaefliIdentityForCurvedSimplices}).

The derivative of the dihedral angles can be also computed from their definition (\ref{ckappa},\ref{dihedral})
\ba
\frac{\partial \theta_{[kl]}}{\partial l_{op}}=-\frac{1}{\sin\theta_{[kl]}}
\frac{1}{\sqrt{ c_{kk} c_{ll}}}
\frac{1}{2} \sum_{h,m}  \left(
c_{kh}c_{ml}+c_{km}c_{hl}-\frac{c_{kl}}{c_{kk}}c_{kh}c_{km} -\frac{c_{kl}}{c_{ll}}c_{lh}c_{lm}
\right) \frac{\partial G_{hm}}{\partial l_{op}}  \, ,
\ea
see also \cite{bdlfss} for simplifications in the Euclidean case.

Finally the  volume of an Euclidean $D$--simplex can be computed by using a generalization of Heron's formula
\begin{eqnarray}
V^2_{(12\ldots
D+1)}\;=\;(-1)^{D+1}\frac{1}{2^D(D!)^2}\;\det\left(\begin{array}{ccccc}0&1&1&\cdots&
1\\1&0&l_{12}^2&\cdots&l_{1D+1}^2\\1&l_{12}^2&0&\cdots&l_{2D+1}^2\\\vdots&
& & \ddots& \\ 1& l_{1D+1}^2 & l_{2D+1}^2&\cdots &
l_{D+1D+1}^2\end{array}\right)
\end{eqnarray}
and one can show that for the derivative of this volume we have
\begin{eqnarray}
\cos \theta_{[ij]}\;=\;\frac{D^2}{V_{(\hat i)}V_{(\hat
j)}}\,\frac{\partial V^2_{(12\ldots
D+1)}}{\partial(l_{ij}^2)}
\;=\;\frac{D^2}{2l_{ij}\,V_{(\hat i)}V_{(\hat j)}}\,\frac{\partial
V^2_{(12\ldots D+1)}}{\partial l_{ij}}  \q .
\end{eqnarray}

\vspace{1cm}

\noindent
{\bf \large Acknowledgements}\\
\noindent It is a pleasure to thank John Barrett, Philipp H\"ohn,
Renate Loll and Ruth Williams for discussions. The research of B.D.
at the University of Utrecht was supported by a Marie Curie
Fellowship of the European Union.

\end{document}